%% file: main.tex
\documentclass[aps,prx,twocolumn,superscriptaddress, nofootinbib, longbibliography]{revtex4-2}
\usepackage{graphicx} 
\usepackage{dcolumn}  
\usepackage{bm}    
\usepackage{amssymb}  
\usepackage{amsfonts, amsmath, amssymb, mathtools}
\usepackage{amsmath}

\usepackage[T1]{fontenc}
\usepackage[utf8]{inputenc}
\usepackage{lipsum}
\usepackage[normalem]{ulem}
\usepackage{commath}
\usepackage{xcolor}
\usepackage{comment}
\usepackage[toc,page]{appendix}
\usepackage{booktabs}
\usepackage[normalem]{ulem}
\usepackage{hyperref}
\usepackage{cleveref}
\usepackage{enumitem}
\RequirePackage{textcase}

\begin{document}

\raggedbottom

\newcommand{\black}[1]{\textcolor{black}{#1}}
\newcommand\myuparrow{\mathord{\uparrow}}
\newcommand\mydownarrow{\mathord{\downarrow}}
\newcommand{\bra}[1]{\langle #1 |}
\newcommand{\ket}[1]{\left| #1 \right\rangle}
\title{Flux-Activated Resonant Control of a Bosonic Quantum Memory}

\author{Fernando Valadares}
\thanks{Contributed equally to this work}
\affiliation{Centre for Quantum Technologies, National University of Singapore, Singapore}
\author{Aleksandr Dorogov}
\thanks{Contributed equally to this work}
\affiliation{Centre for Quantum Technologies, National University of Singapore, Singapore}
\author{Tanjung Krisnanda}
\affiliation{Centre for Quantum Technologies, National University of Singapore, Singapore}
\author{May Chee Loke}
\affiliation{Centre for Quantum Technologies, National University of Singapore, Singapore}
\author{Ni-Ni Huang}
\affiliation{Centre for Quantum Technologies, National University of Singapore, Singapore}
\author{Héctor Calero Mas}
\affiliation{Centre for Quantum Technologies, National University of Singapore, Singapore}
\author{Seungwon Jin}
\affiliation{Centre for Quantum Technologies, National University of Singapore, Singapore}
\author{Pengtao Song}
\thanks{Current address: School of Automation Science and Engineering, Xi’an Jiaotong University, Xi’an 710049, China}
\affiliation{Centre for Quantum Technologies, National University of Singapore, Singapore}
\author{Yvonne Y. Gao}
\email[Corresponding author: ]{yvonne.gao@nus.edu.sg}
\affiliation{Centre for Quantum Technologies, National University of Singapore, Singapore}
\affiliation{Department of Physics, National University of Singapore, Singapore}
\date{\today}

\date{\today}

\begin{abstract}

Universal control of bosonic degrees of freedom provides a hardware-efficient route for quantum information processing with high-dimensional systems. Bosonic circuit quantum electrodynamics (cQED), which leverages auxillary transmons to coherently control long-lived superconducting cavities, is well suited to this goal. However, such systems are traditionally operated in the dispersive regime, where the nearly degenerate cavity transitions prohibit the direct addressability of individual excitation levels of the bosonic mode and increase gate complexity. Here, we achieve direct oscillator control by dynamically accessing the resonant Jaynes–Cummings (JC) interactions, implemented with a hardware that integrates on-chip broadband magnetic flux delivery with a bosonic memory housed in a 3D superconducting cavity with lifetime exceeding 0.5\,ms. We demonstrate deterministic preparation of Fock states and their superpositions within 100s of nanoseconds by directly climbing the JC ladder, and realise efficient arbitrary rotations between \black{a} pair of Fock states. Our resonant control scheme provides an analytical method for manipulating the entire Hilbert space of the bosonic mode at rates \black{potentially} more favourable than traditional dispersive strategies.
\black{This on-demand access to JC interactions serves as a foundational demonstration of an alternative strategy toward realising robust Fock-basis qudits and harnessing the rich dynamics of high-dimensional bosonic systems for quantum information processing.}
\end{abstract}

\maketitle

Effective control over a large Hilbert space is essential for both practical quantum information processing~\cite{shor_scheme_1995, zhou_achieving_2018, huang_learning_2023} and explorations of the rich underlying concepts of quantum physics~\cite{altman_quantum_2021, hofstetter_quantum_2018, abanin_observation_2025}. A promising route towards this goal is to leverage the intrinsically large Hilbert space of quantum harmonic oscillators. Notably, bosonic circuit QED systems use long-lived superconducting cavities coupled to auxillary transmons to realise highly controllable and coherent oscillators. Such devices have demonstrated Fock state preparation up to 100 photons~\cite{deng_heisenberg_2024} and cat states with an average photon number of $256$~\cite{milul_superconducting_2023}. Compared to parallel pursuits using discrete $d$-level systems (qudits)~\cite{lanyon_simplifying_2008, wang_qudits_2020}, such as molecular spins~\cite{rubin2026implementing, moreno_molecular_2018}, optical photons~\cite{wang_experimental_2021, kim_qudit_2024}, trapped ions~\cite{low_control_2025, shi2026efficient}, silicon nuclear spins~\cite{yu_schrodinger_2025} and superconducting artificial atoms~\cite{wang_high_2025, liu_performing_2023, champion_efficient_2025}, encoding information in higher levels of the oscillator \black{in principle} does not introduce additional decoherence channels and is readily compatible with a broad range of existing bosonic quantum error correction schemes~\cite{terhal_towards_2020, grimsmo_quantum_2020}.

\begin{figure}
\centering
\includegraphics{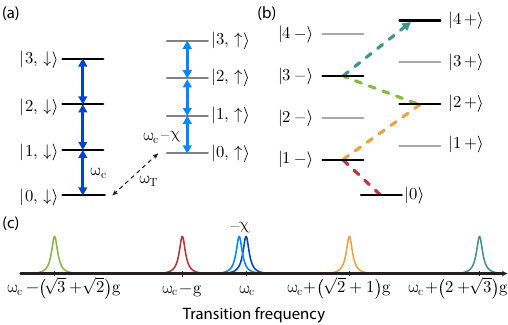}
\caption{\textbf{Comparison of the dispersive and resonant regimes}. \textbf{(a)} The bosonic transitions in the dispersive regime are degenerate within each qubit manifold and are separated by dispersive coupling strength, $\chi$, between the $|\mydownarrow\rangle$ and $|\myuparrow\rangle$ manifolds. \textbf{(b)} This degeneracy is lifted in the resonant JC regime, with level splittings $\propto g \gg \chi$, enabling fast selective excitation of individual bosonic levels. \textbf{(c)} The frequency spectra of both coupling regimes shown on the same axis.}
\label{fig:fig1}
\end{figure}

In cQED systems, universal control of the bosonic mode is typically achieved via dispersive coupling to a nonlinear auxillary element, commonly in the form of a transmon qubit~\cite{blais_circuit_2021}. In this regime, oscillator transitions are close in frequency and not individually addressable, as illustrated (Fig.~\ref{fig:fig1}a). 
Therefore, arbitrary operations on the oscillator state often require driving the transmon and the cavity simultaneously with complex numerically optimised waveforms~\cite{krastanov_universal_2015, heeres_cavity_2015, heeres_implementing_2017}. However, these numerical pulses are susceptible to parasitic dynamics, and their `black-box' nature often complicates effective troubleshooting. Although operations can also be implemented using higher-order parametric processes without numerical optimization~\cite{huang2026universal,huang2026fast, kim_ultracoherent_2025}, they are limited by the achievable coupling strength or drive-induced errors~\cite{verney_structural_2019, lu2026systematic}. Therefore, \black{implementation of an analytical and versatile control method~\cite{strauch_resonant_2012}} that directly addresses individual cavity transitions would fill an important gap for bosonic quantum information processing.

In this work, we use flux-activated resonant Jaynes-Cummings (JC) interactions to access number-selective cavity transitions and implement arbitrary yet analytically simple control of a bosonic memory mode. These dynamics are enabled by adjusting the transmon-cavity detuning on demand via a broadband on-chip flux line. This allows us to rapidly switch the coupled system between a highly nonlinear resonant dynamics with well-resolved JC transitions and the dispersive regime with suppressed cavity photon loss. We protect the coherence of the bosonic memory against flux line leakage by incorporating on-chip low-pass filtering over the frequency band occupied by the cavity and the transmon. As a result, the lifetime of the bosonic mode reaches $T_1 =590\pm40\,\mu\textnormal{s}$ ($Q \sim 2.6\times10^7$) when the transmon is tuned to be dispersively coupled -- more than a two-fold improvement over previous implementations using the magnetic hose~\cite{valadares_ondemand_2024, atanasova_insitu_2025}. We leverage these hardware capabilities to create Fock states and their superpositions by sequentially climbing the JC ladder. Furthermore, we demonstrate state transfer between any two chosen eigenstates of the cavity and realise arbitrary rotations between them using multi-tone pulses that address several JC transitions simultaneously. We present the
first experimental demonstration of using the resonant JC interactions to implement oscillator control, a scheme that can readily be applied to the general class of spin-boson systems, such as opto-acoustic modes coupled to artificial atoms~\cite{bild_schrodinger_2023} or motional degrees of freedom of ions controlled by a spin~\cite{matsos_universal_2025}. 
\black{
Overall, our work showcases the potential of resonant control for quantum information processing using bosonic modes in general qubit-oscillator systems.
}

The resonant control technique explored in this work is based on the theoretical proposal by Strauch~\cite{strauch_resonant_2012}, which adiabatically connects the eigenstates of the resonant JC Hamiltonian to the product eigenbasis of the decoupled circuits. It offers a more efficient approach compared to earlier works~\cite{law_arbitrary_1996,hofheinz2009synthesizing,valadares_ondemand_2024} that rely on sequentially loading single excitations from the spin to the bosonic mode, and provides, in principle, robust universal control over the entire bosonic space.


The dynamics of a spin-boson system on resonance is well approximated within the rotating wave approximation by the JC Hamiltonian~\cite{jaynes_comparison_1963}
\begin{equation}
    \hat{H}_{\textnormal{JC}} = g\left( 
    \hat{a}^\dagger \hat{\sigma}^- + \hat{a}\hat{\sigma}^+\right),
\end{equation}
where $\hat{a}$ ($\hat{a}^\dagger$) is the annihilation (creation) operator of the boson in the form of a superconducting cavity, $\hat{\sigma}^-$ ($\hat{\sigma}^+$) is the lowering (raising) Pauli operator of the spin, realised with a transmon qubit, and $g$ is the coupling strength between them.
The $\hat{H}_{\textnormal{JC}}$ eigenstates are the ground state $|0,\,\downarrow\rangle$ and the $N$th-excited levels
\begin{equation}
\label{eq:jc_eigenstates}
    |N\pm\rangle = \frac{1}{\sqrt{2}}\left( |N,\,\downarrow\rangle \pm |N\mathrm{-}1,\,\uparrow\rangle \right),
\end{equation}
where $|N,s\rangle$ is the product of the cavity Fock state $|N\rangle$ and the transmon state $|s =\, \downarrow,\,\uparrow\rangle$, with eigenfrequencies $\omega_{|N\pm\rangle}=\pm\sqrt{N}g$. This photon-number dependence of $\omega_{|N\pm\rangle}$ leads to the direct addressability of the transitions as illustrated in Fig.~\ref{fig:fig1}(b).

We are particularly interested in the red and blue JC sidebands
\begin{equation}
\label{eq:sidebands}
\begin{aligned}
    |N+\rangle\, \overset{\omega_{N,r}}{\longleftrightarrow}\, |(N\mathrm{+}1)-\rangle, \\
    |N-\rangle\,  \overset{\omega_{N,b}}{\longleftrightarrow} \,|(N\mathrm{+}1)+\rangle,
\end{aligned}
\end{equation}
with transition frequencies $\omega_{N,r}=-(\sqrt{N\mathrm{+}1} + \sqrt{N})g$ and $\omega_{N,b} = (\sqrt{N\mathrm{+}1} + \sqrt{N})g$ relative to the rotating frame, respectively. Transitions $\omega_{1, b}$, $\omega_{2, r}$, and $\omega_{3, b}$ are shown with orange, green, and teal, in Fig.~\ref{fig:fig1}. The sideband transitions, together with ground state excitations $|0,\,\downarrow\rangle\leftrightarrow|1\pm\rangle$, can selectively connect any two eigenstates of the resonant JC Hamiltonian. \black{In the limit of large $N$, the spectral resolution is limited by the frequency separation between the adjacent sidebands $\propto g/\sqrt{N}$ (Fig.~\ref{fig:fig1}c). 
Within the range of photon numbers accessible in typical dispersive controls~\cite{blais_circuit_2021}, the timescale of our protocol is fundamentally more favourable~\cite{sm}.}




To translate this fast control on the JC levels to effective operations on the bosonic mode alone, we simply need to adiabatically decrease the transmon frequency away from the cavity resonance. This transforms the coupled eigenstates in Eq.~\eqref{eq:jc_eigenstates} into products of cavity and transmon eigenstates: $|N+\rangle \to|N,\,\downarrow\rangle$ and $|N-\rangle \to|N\mathrm{-}1,\, \uparrow\rangle$. The red sidebands in Eq.~\eqref{eq:sidebands} map to the transitions $|N,\, \downarrow\rangle \leftrightarrow |N,\, \uparrow\rangle$, similar to the standard transmon excitations in the dispersive regime~\cite{schuster_resolving_2007}. In contrast, the blue sideband, $|N\mathrm{-}1,\,\uparrow\rangle \leftrightarrow |N\mathrm{+}1,\,\downarrow\rangle$, is only directly accessible in the resonant JC dynamics~\cite{strauch_resonant_2012}. Therefore, the on-demand access to the resonant JC dynamics, together with the ability to adiabatically detune the transmon from the oscillator, allows us to orchestrate any desired operations on bosonic mode.

\begin{figure}[!t]
\centering
\includegraphics{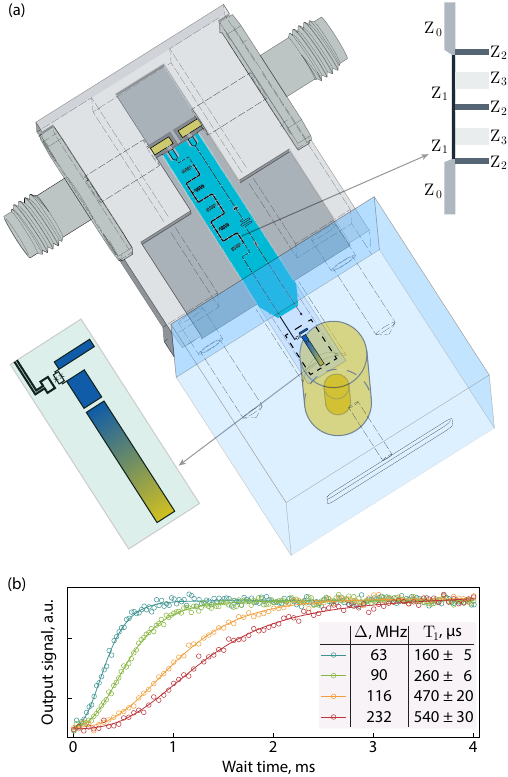}
\caption{\textbf{Architecture for robust flux-delivery in a high-Q bosonic circuit QED device}. \textbf{(a)} High-Q $\lambda/4$ coaxial cavity coupled to an asymmetric SQUID-transmon (shown in bottom-left inset). The flux bias is delivered by a transmission line wirebonded to an SMA port at the opposite end of the chip. The flux line and a readout resonator partially adopt a coplanar waveguide (CPW) architecture. They are equipped with two $3$rd-order Chebyshev low-pass filters in series (schematics shown in top-right inset) and a Purcell filter, respectively, to protect the cavity and the transmon against decay. \textbf{(b)} The cavity $T_1$ is measured while the transmon is parked at a few selected flux points with different values of $\Delta = (\omega_{\textnormal{cav}} - \omega_\textnormal{T})/2\pi$, showing an increase in $T_1$ as the transmon-induced decay is suppressed at larger detunings.}
\label{fig:fig2}
\end{figure}

To leverage this resonant control scheme, we construct a new bosonic circuit QED hardware that can simultaneously provide strong and broadband flux control while protecting the cavity from decay (Fig.~\ref{fig:fig2}a). Our device features a superconducting stub cavity as the long-lived quantum memory coupled to a transmon qubit with a coupling strength $g/2\pi=12.2\pm0.1\,\textnormal{MHz}$. It is combined with a lithographically defined on-chip flux line placed in close proximity to the SQUID loop with area $470\,\mu \textnormal{m}^2$. Flux delivery is highly efficient, featuring a mutual inductance of $0.387\,\Phi_0/\textnormal{mA}$. This allows us to minimise the loop area and hence reduce the impact of flux-noise on the transmon coherence times~\cite{sm}. In this particular device, we measure the highest transmon  $T_1 \approx23\,\mu\textnormal{s} $ and $T_2^{\textnormal{echo}}\approx3.5\,\mu\textnormal{s}$ in the lower sweet spot. The flux line bandwidth reaches at least $1.3\,\textnormal{GHz}$, allowing fast transmon frequency adjustment in the nanosecond timescale. Although we did not calibrate at higher frequencies due to the presence of a 2\,GHz low-pass filter on the coaxial line, our finite-element simulations predict the bandwidth $\sim\textnormal{4}$\,GHz, limited by the on-chip low-pass filter~\cite{sm}.

This CPW-based chip design is carefully optimised to reduce the energy loss of cavity and transmon through the flux line. In comparison to the microstrip flux line shown in a recent work~\cite{li2026cascaded}, our strategy provides a more compact solution that allows convenient integration of additional circuits such as the readout resonator and cascaded filtering on the same chip. We use two $3$rd-order low-pass Chebyshev filters~\cite{pozar2021microwave} with cutoff frequency of 4 GHz in series, schematically depicted in top-right inset of Fig.~\ref{fig:fig2}a. Based on finite element simulations, this structure achieves $98\,\textnormal{dB}$ attenuation at the cavity frequency of $\omega_{\textnormal{cav}}/2\pi = 6.868\,\textnormal{GHz}$ and at least $50\,\textnormal{dB}$ attenuation in the transmon frequency range of $\omega_\textnormal{T}/2\pi=5.894-7.634\,\textnormal{GHz}$. As a result, energy loss through the flux line is heavily suppressed for both quantum memory and transmon. In comparison to the 3D flux-tunable architecture \cite{valadares_ondemand_2024, gargiulo2021fast}, our device achieves enhanced filtering performance and reproducibility with the compact CPW architecture while offering larger flux line bandwidth and longer cavity lifetime.
We verify this by  preparing the cavity in a coherent state and tracking the vacuum state $|0\rangle$ population over time at a few different cavity-transmon detunings as shown in Fig.~\ref{fig:fig2}b. The lifetime reaches a maximum of $T_1 =590\pm40\,\mu\textnormal{s}$ at a detuning of $\Delta = ( \omega_{\textnormal{cav}} - \omega_\textnormal{T} )/2\pi = 260\,\textnormal{MHz}$, at which point the inverse-Purcell decay is suppressed. These photon loss values are comparable to those of typical 3D bosonic superconducting devices without flux integration. More details regarding the transmon and cavity coherence at different flux-biasing points are presented in the supplementary materials.

\begin{figure}[!t]
\centering
\includegraphics{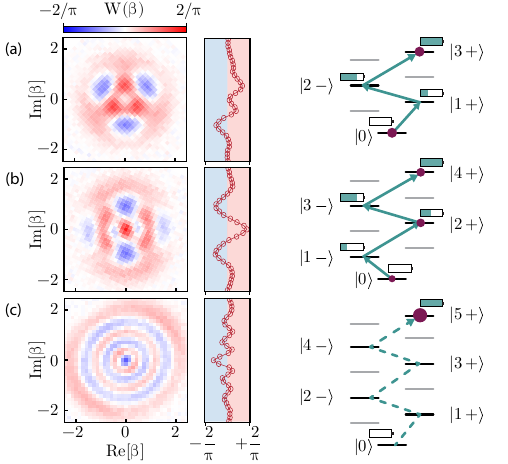}
\caption{\textbf{State preparation in the resonant regime.}
Left panel: Normalised Wigner plots and 1D cuts along $\mathrm{Re}[\beta]=0$ of states created using the flux-activated resonant interactions. Right panel: corresponding transitions activated to climb the JC ladder. \textbf{(a)} State $\tfrac{1}{\sqrt{2}}\left(|0\rangle + |3\rangle\right)$, and \textbf{(b)} state $\tfrac{1}{2}|0\rangle + \tfrac{\sqrt{3}}{\sqrt{8}}\left( i|2\rangle+ |4\rangle \right)$ are created using sequential drives. \textbf{(c)} Fock state $|5\rangle$ created from vacuum by simultaneously driving multiple JC transitions. The progress bars illustrate the steps of state preparation, and the circles indicate the resulting Fock populations.}
\label{fig:fig3}
\end{figure}

Our device combines a highly coherent bosonic mode with fast tunability of a nonlinear auxiliary element, making it an ideal testbed for resonant control protocols. We first showcase the efficacy of this control framework by preparing arbitrary superpositions of Fock states with a set of analytical pulses. For example, we prepare~$\frac{1}{\sqrt{2}}\left(|0\rangle | +|3\rangle \right)$ (Fig.~\ref{fig:fig3}a) using a pulse sequence consisting of a $\pi/2$-pulse from $|0\rangle$ to $|1+\rangle$ followed by two $\pi$-pulses to $|2-\rangle$ and $|3+\rangle$. The frequencies of each transition are determined using standard spectroscopy experiments, and the pulses durations are chosen to maximise the photon-number parity contrast after climbing each step of the JC ladder~\cite{sm}. We calibrate the optimal speed of the adiabatic detuning by first preparing $|1+\rangle$ on resonance, followed by a flux pulse with a smooth, variable-duration ramp until the system is in a strong dispersive regime with $\Delta = 100\,\textnormal{MHz}$. We then directly measure the cavity $|1\rangle$ population using the photon-number-resolved spectroscopy and choose a minimal ramp duration of $200\,\textnormal{ns}$ that maximises the population of $|1\rangle$. 
We further showcase the ease of orchestrating arbitrary multi-component superpositions of Fock states by creating $\tfrac{1}{2}|0\rangle + \tfrac{\sqrt{3}}{\sqrt{8}}\left( i|2\rangle+ |4\rangle \right)$ (Fig.~\ref{fig:fig3}b) using the same method.
The phase of the superposition is implemented by simply changing the phase of the pulse addressing the corresponding transition. This technique can be directly used to create arbitrary logical states superpositions in the binomial encoding~\cite{michael_new_2016, laha_arbitrary_2026}. The total pulse sequence to prepare the state in Fig.~\ref{fig:fig3}a (b) takes $\textnormal{640}$\,ns ($\textnormal{688}$\,ns), and the corresponding fidelity of the final state is estimated as $\textnormal{93}\pm\textnormal{3}\%$ ($\textnormal{89}\pm\textnormal{2}\%$) from density matrix reconstruction~\cite{krisnanda_demonstrating_2025, krisnanda_experimental_2025}, limited by transmon decoherence~\cite{sm}.

As an alternative to sequentially climbing the JC ladder, several transitions can also be activated simultaneously to realise more complex dynamics. For instance, the analog of the spin angular momentum $J_x$ over a subspace of dimension $d$ of the JC Hamiltonian can be constructed by driving each transition with amplitude
\begin{equation}\label{eq:multi-tone_drive}
    A_n = 2\eta_nA  \sqrt{n\left(d-n\right)}, \,\,\,1\leq n<d,
\end{equation}
where A is the overall pulse amplitude and $\eta_n$ is an experimentally calibrated scaling factor of the drive strength associated to each transition~\cite{sm}. The choice of amplitudes \eqref{eq:multi-tone_drive} is such that in the interaction picture the driving Hamiltonian is equivalent to an angular momentum operator for a spin-d/2 system under the rotating wave approximation. The factor $\eta_n$ needs to be calibrated experimentally to account for the variation of the drive line transmission characteristic at different frequencies. Driving $J_x$ for a total time of $\pi/A$ enacts pairwise state transfers between target levels. Here we use this multi-tone drive to create Fock state $|5\rangle$ from the ground state. In principle, the optimal amplitudes of each spectral component can be analytically derived to implement a perfect state transfer~\cite{strauch_resonant_2012, cook1979coherent}. In practice, to improve the fidelity beyond the analytical solution, we allow $A_n$ to assume complex values and obtain the amplitudes using a gradient descent algorithm which maximises the state preparation fidelity for a fixed time~\cite{strauch_resonant_2012, sm}. We present, in Fig.~\ref{fig:fig3}c, the measured Wigner function of a Fock $|5\rangle$ prepared with this method using a $800\,\textnormal{ns}$ pulse. We obtain a state fidelity of $\textnormal{75}\pm\textnormal{2}\%$, limited by both the transmon decoherence and imperfect calibrations of $\eta_n$~\cite{sm}.
We note that the Wigner data are calibrated with the independently measured parity value of the vacuum state and a 2\% uncertainty in this quantity results in $<1$\% error in the reported fidelities.

\begin{figure*}
\centering
\includegraphics{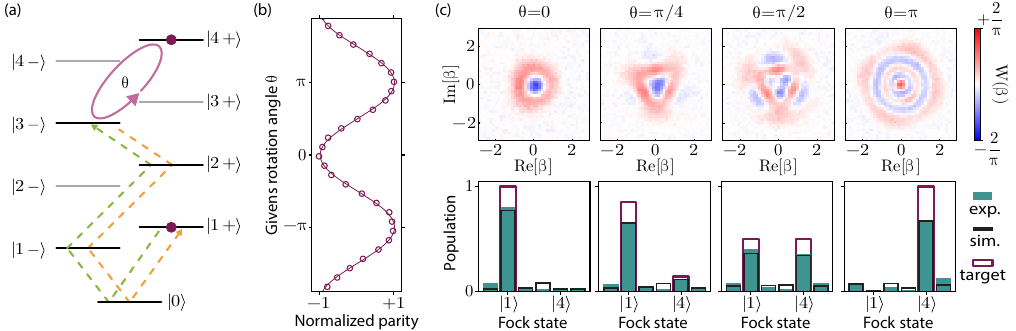}
\caption{\textbf{Givens rotation $\hat{\mathcal{R}}_{1, 4}(\theta)$.}
\textbf{(a)} The $|1+\rangle$ state population is swapped with $|3-\rangle$ state using a $J_x$ pulse, which is then followed by a sideband pulse to $|4+\rangle$ with an effective rotation angle $\theta$ and a second $J_x$ pulse that reverts the first swap. This sequence effectively implements a gate between states $|1+\rangle$ and $|4+\rangle$ that translates into a Givens rotation between Fock states $|1\rangle$ and $|4\rangle$ after adiabatic detuning. \textbf{(b)} Normalised final state parity measured for arbitrary rotation angle $\theta$, showing that the protocol can create arbitrary superpositions of two Fock states. \textbf{(c)} Normalised Wigner function plots and Fock states populations of the cavity state for $\theta = 0, \pi/4, \pi/2$, and $\pi$. The population distributions are obtained from the reconstructed density matrices and consistent with both the ideal state (target) and the outcomes simulated with real experimental parameters (black).}
\label{fig:fig4}
\end{figure*}

Furthermore, we use the multi-tone control to apply arbitrary rotations between two Fock states. This enacts Givens rotations on a qudit in Fock encoding:
$\hat{\mathcal{R}}_{n_A, n_B}(\theta) = \exp [-i \theta/2 (|n_A\rangle \langle n_B| + |n_B\rangle \langle n_A|)]$. Any arbitrary operation on a qudit can be constructed via a combination of $\hat{\mathcal{R}}_{n_A, n_B}(\theta)$~\cite{strauch_resonant_2012,brennen_criteria_2005}, promising a simpler and \black{potentially} faster control compared to the previous work~\cite{mischuck_qudit_2013}.
Our implementation consists of resonantly transferring the population of a level $|n_{A}+\rangle$ to $|\left(n_{B}\mathrm{-}1\right)-\rangle$, and then driving a transition from the latter state to $|n_B+\rangle$ with an effective rotation angle $\theta$ (Fig.~\ref{fig:fig4}a). The sequence is concluded with a second $J_x$ applied in reverse, which swaps $|\left(n_{B}\mathrm{-}1\right)-\rangle$ and $|n_{A}+\rangle$ populations while reverting spectator levels to their original configuration. After the transmon is adiabatically detuned, this operation is equivalent to a Givens rotation between the cavity levels $|n_A\rangle$ and $|n_B\rangle$~\cite{brennen_criteria_2005}. This operation can be naturally extended to perform gates in any chosen subspace, such as the logical states of the binomial encoding, as we demonstrate in~\cite{sm}.
As a concrete example, we apply Givens rotations $\hat{\mathcal{R}}_{1, 4}(\theta)$ between Fock states $|1\rangle$ and $|4\rangle$. The parity of the final state is measured as a function of $\theta$, showing a continuous evolution between a minimum value corresponding to $|1\rangle$ and a maximum corresponding to $|4\rangle$ (Fig.~\ref{fig:fig4}b). For $\theta = 0, \pi/4, \pi/2$ and $\pi$, we measure the full Wigner function of the cavity as shown in the plots in Fig.~\ref{fig:fig4}c. The density matrix reconstruction of the Wigner measurements indicates state fidelities of $\textnormal{80}\pm\textnormal{2}\%$,
$\textnormal{76}\pm\textnormal{2}\%$,
$\textnormal{71}\pm\textnormal{2}\%$,
$\textnormal{68}\pm\textnormal{2}\%$, respectively. 
This is consistent with the numerical simulation with real experimental parameters, including further analysis on process and average gate fidelity~\cite{sm}.

The performance of the Givens rotation and state preparation are limited mainly by transmon pure dephasing $T_{\phi}$ and transmon energy decay $T_1$.
The effect of both error channels on a superposition of JC eigenstates manifest in the fidelity decay $F= 1-\epsilon$, where the infidelity is upper bounded as $\epsilon \lesssim T(1/2T_{\phi}+1/2T_1)$~\cite{sm}.
This gives an approximation for the effective lifetime of the system resulting from transmon decoherence as $T_{\text{eff}}=(1/2T_{\phi}+1/2T_1)^{-1}$.
This analysis allows us to estimate the upper bound on infidelity in our experiments (including the adiabatic detuning process) to be $\approx 20-28\,$\%, which accounts for the majority of error in the fidelities. Our simulation shows that increasing the transmon coherences to $T_1=100\,\mu$s and $T_{\phi}=50\,\mu$s would reduce the upper bound to $\approx2\,$\%.

To enhance the overall fidelity of our resonant control scheme, we can readily adopt the known strategies to mitigate the decoherence of flux-tunable transmons, which is the dominant source of imperfection in this implemenation. For example, the transmon $T_2$ can be increased by reducing the flux sensitivity of the SQUID either by decreasing its area, increasing the asymmetry between the shunting Josephson junctions~\cite{hutchings_tunable_2017, sm}, or using geometries that are inherently more noise-resilient~\cite{braumuller_concentric_2016}. At the same time, flux modulation can be used to engineer AC sweet spots at the cavity frequency to reduce the direct decoherence limit during the operations. Both strategies are supported by the large bandwidth and strong flux delivery of the on-chip line developed in this work. In addition to improving $T_2$, increasing the transmon-cavity coupling strength $g$ leads to more selective sideband transitions and faster adiabatic detuning, both of which can reduce the duration of the protocol and hence increase the fidelity~\cite{sm}. Thus, \black{we expect that} the performance of the resonant operations shown in this work is not fundamentally limited by the control mechanisms or our device architecture, and can be improved by further optimisation of device parameters.

In summary, our work demonstrates the control of a long-lived bosonic quantum memory through number-selective transitions in the JC Hamiltonian. This new capability is enabled by a robust flux-tunable architecture that can effectively switch the system between the dispersive and the resonant regimes. We prepare arbitrary superpositions of Fock states via either sequential or multi-tone sideband driving. Moreover, we use the JC transitions to efficiently implement arbitrary Givens rotations, providing a path for the universal control of qudits encoded in oscillators.
Finally, thanks to the direct access to the JC levels, our protocol is intuitive and calibration-friendly. Moreover, it leverages a frequency selectivity that scales with the coupling factor $g$, allowing \black{potentially} faster control compared to typical dispersive operations and other parametrically activated techniques. 
\black{
Therefore, the resonant control protocol we realised provides effective control of long-lived microwave quantum memories in cQED and opens up new possibilities for quantum information processing with bosonic modes.
}
\\

\noindent\textbf{Acknowledgements} We acknowledge the funding support from the Singapore Ministry of Education (T2EP50222-0017) and The University of Sydney - National University of Singapore 2026 Ignition Grants. F.V., A.D., N.H., H.C.M., S.J. acknowledge the support of the Singapore National Quantum Scholarship Scheme (NQSS). We thank Mr.~Kyle Chu and Mr.~Jean-Samuel Tettekpoe for their technical inputs during the project development. We thank Dr. Yao Lu and Dr. Mustafa Bakr for the valuable discussions during the writing of this manuscript. 

\clearpage 

\onecolumngrid 

\begin{center}
  \textbf{\large Supplementary Materials\\[.4cm]Flux-Activated Resonant Control of a Bosonic Quantum Memory}\\[1cm]
\end{center}

\twocolumngrid

\setcounter{equation}{0}
\setcounter{figure}{0}
\setcounter{table}{0}

\makeatletter
\renewcommand{\theequation}{S\arabic{equation}}
\renewcommand{\thefigure}{S\arabic{figure}}
\renewcommand{\thetable}{\Roman{table}}
\makeatother

\input{supple_contents.tex}

\end{document}

%% file: supple_contents.tex
\raggedbottom

\section{Device Design}

The standard bosonic cQED architecture consists of a high-Q bosonic mode hosted in a 3D superconducting cavity with convenient control and measurement tools provided by an on-chip ancillary transmon and a planar low-Q readout resonator. Our device extends this design by integrating fast-flux tunability without compromising long cavity lifetimes. 

\begin{figure*}
\centering
\includegraphics{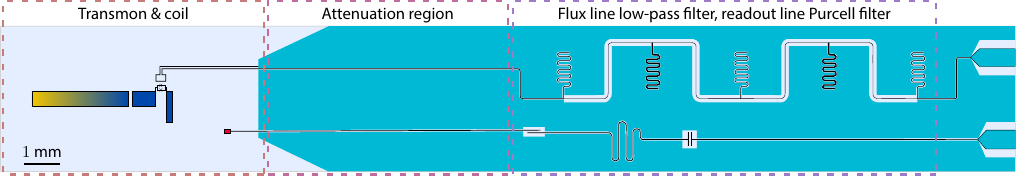}
\caption{\textbf{Chip schematic}. It features ancilla DC-SQUID transmon, low-Q readout resonator with Purcell filter, and on-chip flux line with microwave low-pass filter}.
\label{fig:sfig_chip}
\end{figure*}

These flux control features are achieved by combining coplanar waveguide (CPW) and microstrip line architectures. We show the details of the circuit in Fig.~\ref{fig:sfig_chip}. The tunable transmon is designed in a microstrip geometry, allowing it to have a large dipole moment that couples strongly to the cavity. The SQUID loop of the transmon is inductively coupled to an on-chip coil biased by a CPW flux line. The flux line is designed with an integrated microwave low-pass filter that mitigates cavity and transmon photon loss. 

In this section, we explain the technical specifications and design considerations of the tunable transmon and the flux line control.

\subsection{Tunable transmon with asymmetric SQUID}

Standard transmon~\cite{koch2007charge} qubits feature nonlinear inductance implemented as a Josephson junction, leading to anharmonic oscillator spectrum where one can isolate two-dimensional computational subspace with the transition frequency $\omega_\textnormal{T}$ given by

\begin{equation}\label{eq:single_junction}
    \hbar \omega_{\textnormal{T}} = \sqrt{8E_C E_J} - E_C,
\end{equation}
where $E_J$ and $E_C$ stand for Josephson and capacitive energies, respectively.

\begin{figure}[h]
\centering
\includegraphics{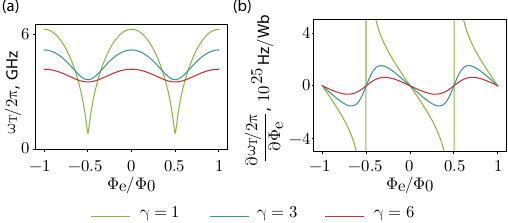}
\caption{\textbf{Reduced noise sensitivity of asymmetric SQUID transmons.} \textbf{(a)} Frequency response for variable junction asymmetry factors $\gamma$. \textbf{(b)} Calculated transmon frequency sensitivity to flux noise as a function of flux bias.}
\label{fig:sfig_asymmetric}
\end{figure}

In order to make such a qubit tunable, single Josephson junction should be replaced by a loop with two junctions in parallel, forming a DC-SQUID. Magnetic flux $\Phi_e$ threaded through the SQUID loop effectively changes the value of the inductance and, therefore, the qubit frequency, according to

\begin{equation}\label{eq:sym_squid}
    \hbar \omega_\textnormal{T} = \sqrt{8E_C\cdot 2 E_J |\cos{\varphi_e}|} - E_C,
\end{equation}
where $E_J$ is the Josephson energy of a single junction, $\varphi_e = \pi \frac{\Phi_e}{\Phi_0}$ the normalised flux , and $\Phi_0$ the magnetic flux quantum. Typical spectrum of $\omega_\textnormal{T}$ as a function of the applied magnetic flux is shown in Fig.~\ref{fig:sfig_asymmetric}(a).

In such tunable devices, noise in the DC bias or in the magnetic environment leads to the fluctuations of the magnetic flux. This, in turn, varies the transmon frequency and leads to significant limitations on the $T_2$ lifetime. Thus, it is beneficial to operate at the sweetspots, where the qubit is first-order insensitive to the flux noise. At other bias points, the qubit suffers rapidly increasing dephasing. 

This sensitivity to the flux noise can be quantified as $\frac{\partial \omega_\textnormal{T}}{\partial \Phi_e}$, i.e. how much the qubit frequency reacts to the fluctuations of the magnetic flux due to the noise. From Eq.~\eqref{eq:sym_squid},

\begin{equation}\label{eq:symmetric_sensitivity}
    \frac{\partial \omega_\textnormal{T}}{\partial \Phi_e} = -\frac{\pi^2}{\Phi_0} \sqrt{2\frac{E_C \cdot 2 E_{J}}{h^2}} \frac{\sin{2\varphi_e}}{\cos^{\frac{3}{2}}{\varphi_e}}.
\end{equation}

The sensitivity to the flux noise grows rapidly outside of the sweetspots due to the strong dependence of frequency on the applied flux bias. Here, we intentionally suppress this dependency by making the two junctions asymmetric, i.e. having different Josephson energies~\cite{hutchings2017tunable}. The extent of asymmetry is expressed as 

\begin{equation}
    \gamma = \frac{E_{J2}}{E_{J1}},
\end{equation}
where $E_{J1}$ and $E_{J2}$ are Josephson energies of the junctions and $\gamma > 1$ without loss of generality. 

Naturally, the spectrum of the asymmetric transmon will also depend on the parameter $\gamma$:
\begin{equation}\label{eq:asymmetric_squid}
    \hbar \omega_\textnormal{T} = \sqrt{8E_C\cdot E_{J1}\sqrt{\gamma^2 + 2 \gamma \cos{2\varphi_e} + 1}} - E_C.
\end{equation}
We calculate the flux noise sensitivity analogously to Eq.~\eqref{eq:symmetric_sensitivity} using Eq.~\eqref{eq:asymmetric_squid} and obtain
\begin{equation}\label{eq:asymmetric_sensitivity}
    \frac{\partial \omega_\textnormal{T}}{\partial \Phi_e} = -\frac{\pi^2}{\Phi_0} \sqrt{\frac{2E_C E_{J1}}{h^2}} \frac{4\gamma \sin{2\varphi_e}}{\big( \gamma^2 + 2\gamma\cos{2\varphi_e} + 1 \big)^{\frac{3}{4}}}.
\end{equation}

We compute flux sensitivies using Eq.~\eqref{eq:symmetric_sensitivity} and Eq.~\eqref{eq:asymmetric_sensitivity} in Fig.~\ref{fig:sfig_asymmetric}(b). One can observe that more asymmetric transmons are less sensitive to the flux noise at the expense of a smaller frequency range. Therefore, for this particular device, we designed the SQUID to be asymmetric with a parameter $\gamma = 4.02$. The resulting frequency range is 5.894 - 7.634 GHz, as shown in Fig.~\ref{sfig:filtering}(a).

It is worth mentioning that Eq.~\eqref{eq:sym_squid} and further analysis, strictly speaking, is only valid for the applied biasing DC current. In case of an AC flux drive, a more rigorous analysis~\cite{you2019circuit, lu2026systematic} must be applied to ensure an effective differential drive on the SQUID. The use of an asymmetric SQUID makes this optimisation more complicated. Although it is outside of the scope of this work, we are actively exploring alternative strategies that can balance the robustness to flux noise with the effective suppression of the common mode drive for future works that require clean AC modulation of the qubit frequency.

\subsection{Flux line \& on-chip microwave low-pass filter}

The integration of on-demand adjustment of the transmon frequency without reducing the cavity coherence times had been an outstanding challenge for the bosonic cQED community. This feature is crucial as it enables powerful quantum information processing protocols~\cite{strauch2012all, terhal2016encoding}, as well as new opportunities for quantum simulation~\cite{joshi2016quantum, wang2019simulating} and quantum machine learning~\cite{henderson2024quantum, tolstobrov2024hybrid}. However, initial attempts of the implementation of the broadband on-chip flux line encountered an issue of compromising cavity coherence~\cite{reed2013entanglement}. In recent efforts, alternative ways of the flux delivery to the SQUID loop of the tunable ancilla have been explored ~\cite{gargiulo2021fast, valadares2024demand, atanasova_insitu_2025}. However, these implementations tend to be difficult to manufacture reproducibly and offer limited bandwidth ($\sim$ 100 MHz). 

In this work, we demonstrate the device that combines an on-chip flux line with both effective DC and AC flux delivery with a long-lived cavity. That is achieved by implementing heavy filtering of the flux line: we incorporate compact, third order Chebyshev low-pass filter~\cite{pozar2021microwave} with cutoff frequency of 4 GHz on the same chip. To ensure effective filtering performance and protection of the cavity mode, we concatenate two such filters, with the stub $Z_2$ shared between the two, as depicted schematically in Fig.~\ref{fig:sfig_chip}. We leverage the CPW architecture, which allows to reach necessary impedances in much more compact design compared to the microstrip analogue~\cite{li2026cascaded} and to fit both filtering stages on the 4.1 mm-wide chip. Moreover, our CPW architecture also suppresses cross-talk between the flux and readout modes hosted on the same chip. 

The impedance values of the series inductances and stub capacitances used in our filter are listed in Tab.~\ref{tab:impedances}, along with the corresponding CPW line dimensions.

\begin{table}[h!]
    \centering
    \begin{tabular}{|c|c|c|c|}
         \hline
         Section & Impedance, $\Omega$ & Conductor width, $\mu$m & Gap width, $\mu$m \\
         \hline \hline
         $Z_0$ & 50 & 11 & 5 \\
         \hline
         $Z_1$ & 130 & 5 & 100 \\
         \hline
         $Z_2$ & 81 & 5 & 13 \\
         \hline
         $Z_3$ & 46 & 15 & 5 \\
         \hline
    \end{tabular}
    \caption{\textbf{Impedances of the 3-rd order Chebyshev low-pass filter}. Filter diagram is shown in Fig. 2 (a) of the main text.}
    \label{tab:impedances}
\end{table}

\begin{figure}[h!]
\centering
\includegraphics{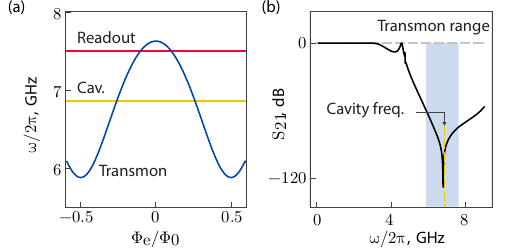}
\caption{\textbf{Filtering and spectrum}. \textbf{(a)} A spectrum of the transmon with asymmetric SQUID. Cavity and readout resonators frequency are shown. \textbf{(b)} $S_{21}$ parameter of the cascaded low-pass 3-rd order Chebyshev filter.}
\label{sfig:filtering}
\end{figure}

The simulated response of the on-chip low-pass filter is shown in Fig.~\ref{sfig:filtering} (b). The attenuation at the cavity frequency (6.868 GHz) is 98\,dB. For the entire transmon frequency band, the attenuation is at least 50\,dB.

\begin{figure}[h!]
\centering
\includegraphics{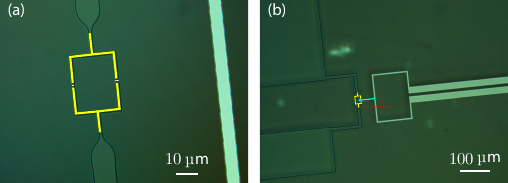}
\caption{\textbf{Optical images of the device after electron beam lithography}. \textbf{(a)} SQUID loop (yellow) with asymmetric junctions (white). \textbf{(b)} Flux line coil, biasing the transmon.}
\label{sfig:microscope}
\end{figure}

In this device, the SQUID area is 470\,$\mu$m$^2$ (Fig.~\ref{sfig:microscope}(a)), and the flux line coil loop is placed as close as 58\,$\mu$m away from it to maximise the coupling (Fig.~\ref{sfig:microscope}(b)). The mutual inductance of $L_{DC} = 0.387\,\Phi_0$/mA is measured from the transmon frequency perodicity with a DC bias. This allows us to reach full tunability (requires $\Phi_e = \frac{1}{2} \Phi_0$) with only 1.292\,mA DC bias, which does not cause any measurable heating of the transmon.

Furthermore, we incorporate a long constriction in the volume where the chip is housed, which we dub the "attenuation region". It considerably increases the cutoff frequency of the waveguide structure, enhancing the suppression of the cavity field traveling outside the ground plane. Hence, the cavity field can only propagate to the drive ports through the CPW mode, which is heavily filtered. 

\subsection{Readout and Purcell filter}

Another advantage of the compactness of the CPW architecture, realised with the hybrid chip design, is its simplicity in integration of the readout resonator and the associated bandpass Purcell filter on the same chip. Our Purcell filter, which is shown in Fig.~\ref{fig:sfig_chip}, is strongly coupled capacitively to both the readout resonator and the feed line. With this design, our readout resonator, which we place at 7.51 GHz, has coupling quality factor of $Q \approx$ 700 without introducing any notable limitations on the transmon $Q$ factor.

\section{Device Fabrication}

The device used in this study has considerably larger footprint compared to typical transmon chips used in bosonic cQED devices due to the ground plane and the flux line with low-pass microwave filter. For practical purposes, we only use EBL for patterning the transmon, while photolithography was used for all the other parts of the chip, as summarised in the Tab.~\ref{tab:fabrication_steps}.
\begin{table*}[htp!]
    \centering
    \begin{tabular}{|c|c|c|c|}
         \hline
         \textbf{\# of lithography step}& \textbf{Exposure method} & \textbf{Material} & \textbf{Written feature} \\
         \hline \hline
         1 & EBL & Nb & Alignment markers \\
         \hline
         2 & Photolithography & Al & Readout line, flux line, ground plane \\
         \hline
         3 & EBL & Al &  Transmon with the SQUID loop\\
         \hline
    \end{tabular}
    \caption{\textbf{Summary of the main fabrication steps}. A breakdown of the processes used in fabricating respective on-chip features.}
    \label{tab:fabrication_steps}
\end{table*}


An optical microscope image of the SQUID right after EBL exposure is shown in Fig.~\ref{sfig:microscope}(a).

\subsubsection*{Step 1: alignment markers}

The fabrication of the device begins with the cleaning of a 2-inch C-plane sapphire wafer by sonicating in N-Methyl-2-pyrrolidone (NMP) and methanol for 4 minutes each, followed by baking at 200$^\circ$C for 5 minutes to remove any residual chemicals molecules. Following this, the wafer is cooled down on a heat sink for 5 minutes. After that we spincoat two layers of the EBL resists: 550\,nm of MMA (8.5) MAA EL13 and 250\,nm of 950k PMMA A4. Each of the spincoating rounds is followed by baking for 5 minutes at 200$^\circ$C. Next, we sputter an anti-charging gold layer of $\sim$10 nm thickness. We then expose the alignment markers using Raith EBPG 5200+ EBL system with the dose 750 $\mu$C/cm$^2$. Subsequently, we etch gold using KI gold etchant and develop the exposed pattern with 3:1 IPA:H2O at 6$^\circ$C for 2 minutes.

To clean leftover resist before the metal deposition, we apply etching for 60 seconds in Ar ($85\,\%$) -- O$_2$ ($15\,\%$) environment and evaporate 115\,nm of Nb using Angstrom Engineering double-angle evaporation system. We perform lift-off via soaking the wafer in NMP at 90$^\circ$C for 3 hours and then sonicating in NMP, acetone and methanol for 3 minutes each.

\subsubsection*{Step 2: readout line, flux line, ground plane}

We start by spincoating $\sim$ 1\,$\mu$m of AZ1512 photoresist and baking for 1 minute at 100$^\circ$C. Once this is completed, we expose the large features with 405\,nm laser in Durham Magneto Optics MicroWriter ML3 Pro. Following this, we develop the exposed resist with MF-319 developer for 40 seconds and DI water for 60 seconds and evaporate 100\,nm of Al. Etching for 30 seconds and the evaporation itself are followed by capping in a 20 mbar O$_2$ environment for 20 minutes. To lift-off, we soak the wafer in NMP at 80$^\circ$C for 4 hours and sonicate in NMP, acetone and methanol 3 minutes each.

\subsubsection*{Step 3: transmon with SQUID}

Finally, we perform standard electron beam lithography to make the trasmon qubit. The junctions are fabricated using the bridge-free technique, presented in~\cite{lecocq2011junction}. We then perform double-angle evaporation by first etching the resist at $\pm$20$^\circ$ angles for 15 seconds each. Subsequently, we deposit two layers of Al with thicknesses of 20\,nm and 40\,nm, separated by oxidation for 20 minutes in 20 mbar O$_2$ environment. We finish the evaporation by capping in 20\,mbar O$_2$ environment for 20 minutes.





\section{Device assembly and wirebonding}

\subsection{Package}

Conventional bosonic cQED architecture usually uses a monolith package, made of high-purity aluminium, to minimise seam loss~\cite{copetudo2024shaping, pan2025realization}. The chip hosting the ancillary elements is inserted into a waveguide to couple capacitively to the cavity mode and the other end of the chip is held by a clamp, which is, in turn, attached to the main package. However, this approach is not suitable for the device we use here. The on-chip ground plane and requirement for wirebonding necessitate a seam placed near the cavity wall. 
\begin{figure}[h!]
\centering
\includegraphics{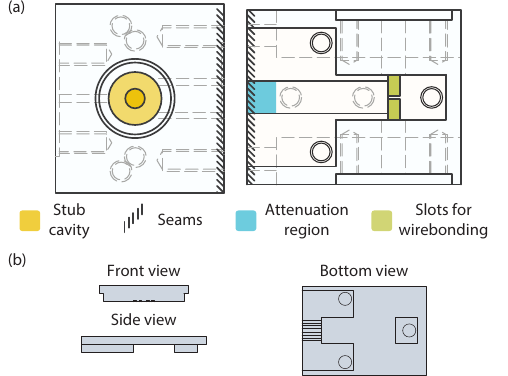}
\caption{\textbf{Design of the superconducting package}. \textbf{(a)} Top view of the main body and the bottom clamp. \textbf{(b)} Design of the top clamp with engineered constrictions.}
\label{sfig:package}
\end{figure}


Hence, our superconducting package was carefully re-designed to accommodate these requirements. Unlike conventional architecture, our package consists of three separate parts. The main body contains the cavity and the circular waveguide, where the transmon is inserted (Fig.~\ref{sfig:package}(a)). This part is made of 4N6 aluminium to ensure the long lifetime of the coaxial $\lambda/4$-cavity. 

The clamping structure consists of two pieces, the bottom and top clamps. As we expect the high-Q cavity mode field to be weak in the clamps volume, these parts are made of 6061 aluminium alloy to allow more precise machining and mechanical robustness.

\begin{figure}[h!]
\centering
\includegraphics{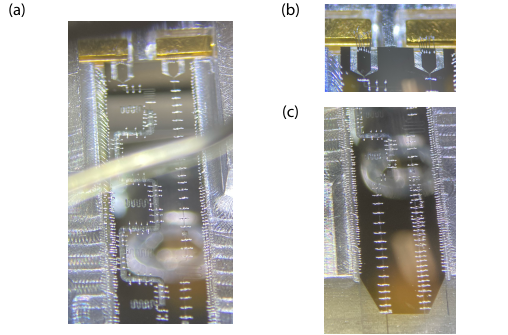}
\caption{\textbf{Wirebonding}. \textbf{(a)} Image of the chip on the bottom clamp after wirebonding. \textbf{(b)} Connection of the on-chip lines to th SMA ports. \textbf{(c)} Bond wires and galvanic connection between the on-chip ground plane and the aluminium bottom clamp are also realised with wirebonding.}
\label{sfig:wirebonding}
\end{figure}

The top clamp functions as a lid that covers the chip and shields it from the environment. Importantly, it features a 4mm-long extrusion (Fig.~\ref{sfig:package}(b)), which blocks the cavity field from propagating towards SMA ports outside the heavily filtered CPW line.

The bottom clamp features a 0.5 mm-deep pocket, where the chip is located (Fig.~2(a), main text). We use wirebonding to connect the on-chip ground to the surface of the clamp surrounding it (Fig.~\ref{sfig:wirebonding}(a),(c)). The package is then mounted on the MXC flange of the dilution refrigerator, effectively connecting on-chip ground plane to the common fridge ground. Similarly, we use wirebonding to galvanically connect the flux and readout line terminals to flat beryllium-copper pins, that are connected to SMA ports (Fig.~\ref{sfig:wirebonding}(b)). The drive signals are supplied to the SMA ports through the coaxial lines of the dilution refrigerator.

\subsubsection*{Seam losses estimation}

To avoid deterioration of the cavity lifetime, our design places the seam as far as possible from the cavity. In order to estimate cavity Q-factor due to the seam loss, we integrated the cavity field over a 0.4 mm-wide ribbon along the seam. Even though we don't have exact data on the loss tangent of the seam, we could use that method to optimise the design based on the relative changes of the cavity Q. 

Distancing the seam comes at the cost of increasing the separation between the cavity and the transmon, which could lead to decreasing their coupling strength $g$. In order to mitigate this effect, we implemented a microstrip structure to extend the transmon field and recover some of the capacitive coupling, similar to the design used in Ref.~\cite{valadares2024demand}.


\subsection{Assembly}

We begin the assembly by putting a small amount of the Apiezon N cryogenic vacuum grease on the surface of the bottom clamp. We then place the chip on top and press slightly with a wooden part of the cotton swab. The grease is only present at the back part of the chip, far from the cavity.

Next, the top clamp is placed on top of the chip and fastened with 3 screws. At this point, the chip is extruding out of the clamp slightly. Finally, the cavity body and the clamping pieces are attached together with the two screws located at the bottom of the clamp.

To improve the contact at the seam in our package, we put indium wire in between the cavity and the clamping parts. The quality of this joint can be further improved with welding~\cite{milul2023superconducting} or more advanced surface treatment~\cite{krayzman2022thin}.

Before mounting the device on the MXC flange of the dilution refrigerator, we cover all the seams between the parts with aluminium tape to ensure that the package is light-tight. 



\section{Device parameters and flux-dependent coherences}\label{ssec:coherences}

\subsection{Device parameters}

\black{In the table below we present the values of the device parameters. The lab-frame resonant Jaynes-Cummings Hamiltonian includes only 3 fixed parameters: the resonant frequency $\omega_c/2\pi$, the coupling strength $g/2\pi$, and the transmon anharmonicity $K / 2\pi$: }

\black{
\begin{equation}\label{eq:JC_Hamiltonian_lab_frame}
    \hat{H}_\textnormal{JC}^\textnormal{lab frame} = \omega_c \hat{a}^\dagger \hat{a} + \omega_c \hat{b}^\dagger \hat{b} + \frac{1}{2} K \hat{b}^\dagger \hat{b}^\dagger \hat{b} \hat{b} + g (\hat{a}^\dagger \hat{b} + \hat{a} \hat{b}^\dagger),
\end{equation}}

\black{where $\hat{a}$ and $
\hat{b}$ are cavity and transmon annihilation operators, respectively, and we set $\hbar$ = 1.}

\black{The parameters below are extracted from the values of the eigenenergies of the JC eigenstates, as listed in the Table \ref{table:anharmonic_eigenenergies}. All the other parameters, such as system lifetimes and the cross-Kerr in the dispersive regime (dispersive shift) are flux-dependent. }

\begin{table}[htp!]
    \begin{center}
    \begin{tabular}{||c|c||} 
     \hline
     JC parameter& Value, MHz\\ [0.5ex] 
     \hline
     $\omega_c/2\pi$ & 6868.737\\
     \hline
     $g/2\pi$ & 12.088\\
     \hline
     $K/2\pi$ & $-$206.020\\ 
     \hline
    \end{tabular}
    \caption{\black{Resonant Jaynes-Cummings Hamiltonian parameters (resonant frequency, coupling strength, transmon anharmonicity), extracted from the values of JC eigenenergies from the Table \ref{table:anharmonic_eigenenergies}.}}
    \label{tab_supple:device_parameters}
    \end{center}
\end{table}

\subsection{Qubit lifetime}

\black{Lifetimes of the ancillary qubit and the cavity at different flux bias points are presented in the Tables \ref{tab_supple:2025}, \ref{tab_supple:2026}, \ref{tab_supple:2026_cav}.}

\begin{table}[htp!]
    \begin{center}
    \begin{tabular}{||c| c| c| c| c||} 
     \hline
     Applied flux, $\Phi_0$ & Frequency, GHz & $T_1$, $\mu$s & $T_2$, $\mu$s & $T_2^\textnormal{echo}$, $\mu$s \\ [0.5ex] 
     \hline
     0.20 & 7.1453 & 4.23 & 1.12 & -- \\
     \hline
     0.21 & 7.0981 & -- & 1.53 & -- \\
     \hline
     0.26 & 6.8701 & 8.7 & 0.96 & -- \\ 
     \hline
     0.50 & 5.9088 & 23.0 & -- & 3.5 \\
     \hline
    \end{tabular}
    \caption{\black{Qubit coherences as a function of flux. Data collected in late 2025, in the same cooldown as the rest of the experimental results.}}
    \label{tab_supple:2025}
    \end{center}
\end{table}

\begin{table}[htp!]
    \begin{center}
    \begin{tabular}{||c| c| c| c| c||} 
     \hline
     Applied flux, $\Phi_0$ & Frequency, GHz & $T_1$, $\mu$s & $T_2$, $\mu$s & $T_2^\textnormal{echo}$, $\mu$s \\ [0.5ex]
     \hline
     0.01 & 7.150 & 0.47 & 0.62 & 0.75 \\
     \hline
     0.18 & 6.785 & 6.1 & 0.32 & -- \\
     \hline
     0.20 & 6.709 & 3.3 & 0.43 & 0.67 \\
     \hline
     0.24 & 6.549 & 2.6 & 0.35 & -- \\
     \hline
     0.33 & 6.127 & 7.6 & 0.28 & -- \\
     \hline
     0.50 & 5.567 & 15.9 & 1.35 & 1.59 \\ 
     \hline
    \end{tabular}
    \caption{\black{Qubit coherences as a function of flux. Data collected in June and July 2026 with the same device. However, the device has been exposed to atmospheric pressure for about 9 months, which results in some degradation of both the cavity and qubit properties.}} 
    \label{tab_supple:2026}
    \end{center}
\end{table}

\black{The data is consistent with the Q-factors extracted from the electromagnetic simulations of the device, though the 2026 data features lower bare lifetimes due to the aging of the device. }
    
\black{Thus, the qubit $T_1$ has minimum values near the upper sweetspot, where it is the most coupled to the Purcell filter and the readout resonator modes that are intentionally made lossy to facilitate faster qubit readout. Maximum qubit $T_1$ was observed in the lower sweetspot, where it is the most protected by the bandpass Purcell filter~\cite{krantz2019quantum} and has the largest detuning from the readout resonator. Obviously, qubit $T_1$ also shows increase near the cavity frequency when the qubit hybridises with the high-Q cavity mode, e.g. the points $\Phi = 0.18\, \Phi_0$ and $\Phi = 0.20\, \Phi_0$ in the Table \ref{tab_supple:2026}.}

\black{The frequency dependence of the qubit $T_1$ indicates that its energy loss is dominated by Purcell decay through the readout line, while not being affected by the on-chip flux line due to incorporation of the ''attenuation region'' and the low-pass filter (see Fig. \ref{fig:sfig_chip}).}

\black{Qubit $T_2$ also peaks in the lower sweetspot, where the qubit is first-order insensitive to the flux noise, while it is less coupled to the other lossy (and ''hot'' in 2026) modes than in the upper sweetspot.  The qubit $T_2$ values outside of the sweetspots don't show any obvious trend due to high asymmetry in the Josephson energy between the two junctions of the SQUID. Although \cite{hutchings2017tunable} predicts slight dependence of dephasing rate on the applied flux outside of sweetspots, it is hard to catch with only a few points measured. On top of that, lack of the trend might serve as an evidence of the qubit $T_2$ being limited by coupling to two-level systems.}

\subsection{Cavity $T_1$}


\black{We observe a general trend where the cavity $T_1$ increases with larger cavity-transmon detuning. This behavior is expected from the inverse-Purcell effect~\cite{reagor2016quantum}, which accounts for the cavity's single-photon loss rate induced by coupling to the lossy transmon:}

\black{
\begin{equation}\label{eq:inverse_Purcell}
    \kappa_{\textnormal{IP}} \approx (g^2 / (2\pi\Delta)^2) \times \gamma_1,
\end{equation} }

\black{where $g$ is the qubit-cavity coupling strength, $\Delta$ is the detuning, and $\gamma_1$ is the transmon energy loss rate. The isolated reduction in $T_1$ at $\Phi = 0.24\Phi_0$ is a deviation from this trend, attributed to coupling with a two-level system (TLS) — shown in Fig.~\ref{sfig:cavity_T2_coherent_decay} (c) — detected concurrently with this measurement.}

\begin{table}[ht]
    \centering
    \begin{tabular}{||c| c| c| c||} 
     \hline
     Applied flux, $\Phi_0$ & $\Delta$, GHz & Cav $T_1$, $\mu$s & Cav $T_2$, $\mu$s \\ [0.5ex] 
     \hline
     0.50 & 1.3 & 350 & 230 \\
     \hline
     0.24 & 0.32 & 135 & 12 \\
     \hline
     0.20 & 0.16 & 150 & -- \\
     \hline
     0.18 & 0.084 & 120 &  -- \\
     \hline
     0.01 & $-$0.28 & 160 & -- \\
     \hline
    \end{tabular}
    \caption{\black{Data collected in June and July 2026 with the same device. }}
    \label{tab_supple:2026_cav}
\end{table}

\black{As the qubit itself is limited by the Purcell effect due to coupling to the lossy readout and Purcell filter modes, analyical modelling and numerical fit to define the bare decay rates are complicated and require larger set of data on qubit and cavity $T_1$ lifetime. However, the qualitative conclusion based on the Tables \ref{tab_supple:2025}, \ref{tab_supple:2026}, \ref{tab_supple:2026_cav} and the Fig. 2 (b) of the main text can be made: the system modes are limited by the cascaded Purcell effect, as the qubit $T_1$ is limited due to coupling to the readout line, and the cavity $T_1$ is limited due to coupling to the qubit.}

\subsection{Cavity dephasing}

\black{Next, let us consider cavity dephasing rate. In the current system, such measurements are rather challenging due to the short transmon coherence time and the inability to perform number-selective qubit rotations that are conventionally used in cavity $T_2$ measurements~\cite{reagor2016quantum}. As such, we designed a few new experiments to estimate this timescale. Although we are not able to probe the dephasing as a function of the flux over a broad range due to the specific technical challenges imposed by the current device parameters, we have selected two bias points to evaluate the representative dephasing times in the strong and weak dispersive regimes, respectively. A complete, high-quality flux-dependent cavity $T_2$ map on a well-thermalised, undegraded device is important, and we plan to carefully investigate dephasing mechanism in the upcoming study.}

\begin{figure*}[htp]
    \centering
    \includegraphics{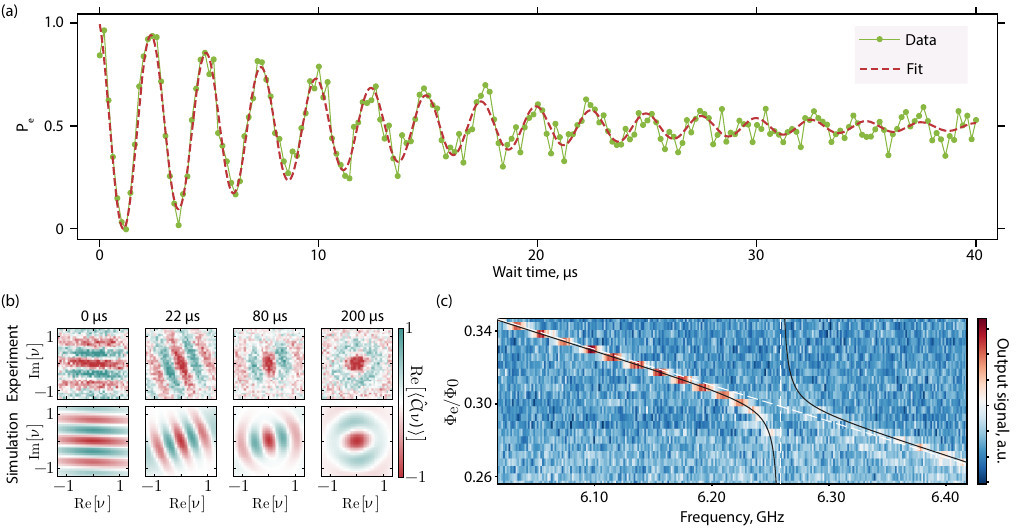}
    \caption{\black{\textbf{Cavity dephasing}. Note that this set of measurements were carried out using the original device but 9 months after the results in the main text were obtained. Both the quality factors and thermalisation of the components suffered some noticeable degradation during this idling time at atmospheric pressure. \textbf{(a)} Cavity $T_2$ experiment in the strong dispersive regime. The value of cavity $T_2$ determined from the fit is $\approx$ 12 $\mu$s. \textbf{(b)} Comparison of the real part of the measured characteristic function to the numerical simulation of the master equation with cavity decoherence for variable decay time of the initial coherent state $\ket{\alpha\approx3}$ in the weak dispersive regime. The value of the cavity $T_2$ obtained from the comparison is $\sim$ 230 $\mu$s. \textbf{(c)} An example of an avoided crossing of the qubit and the two-level system. The solid black line is a fit, from which we reconstruct the coupling constant $g/2\pi \approx$ 29 MHz. The dashed white lines are the guides for the eye, indicating the frequency of the TLS and the unperturbed frequency curve of the qubit.}}.
    \label{sfig:cavity_T2_coherent_decay}
\end{figure*}
    
\black{The experiment to determine the cavity dephasing rate in the strong dispersive regime ($\chi/2\pi \approx$ 0.50 MHz, $\Delta \approx$ 0.32GHz) was designed to resemble the qubit $T_2$ measurement using the Ramsey interferometery~\cite{krantz2019quantum}. First, we prepare the qubit in the state ($\ket{g}+\ket{e}$)/$\sqrt{2}$ by applying a $\pi/2$ pulse. Then we bring the qubit-cavity system on resonance non-adiabatically (in under 16 ns $\ll 2\pi/g$), which projects the initial state onto the superposition of the JC states according to $\ket{0,\, \uparrow} \to \big(\ket{1,\, +} - \ket{1,\, -} \big)/\sqrt{2}$ and $\ket{0,\, \downarrow} \to \ket{0}$. As the eigenenergies are non-degenerate, this superposition will evolve in time as}

\black{
\begin{equation}\label{eq:jc_evolution}
        \ket{\psi(t)} = \big(\ket{1,\, +} \textnormal{e}^{-\imath 2gt} - \ket{1,\, -} \big)\textnormal{e}^{-\imath (\omega_\textnormal{cav} - 2g)t} + \ket{0}.
    \end{equation}}

\black{After holding the system on resonance for $t_1 = \pi/2g$ we non-adiabatically detune the qubit from the resonant regime that projects the system state to $\ket{0,\, \downarrow} - \ket{1, \, \downarrow}\textnormal{e}^{-\imath (\omega_\textnormal{cav} - 2g)t_1}$, effectively creating a superposition of the Fock states $\ket{0}$ and $\ket{1}$ in the cavity while the qubit is in its ground state. Then we wait for a variable time $\tau$ and bring the system back on resonance for the same time $t_1$, effectively swapping the superposition of the Fock states in the cavity to a superposition of the qubit ground and exciting states. Then we apply a $\pi/2$ pulse on the qubit with the phase $\delta = ((\omega_\textnormal{cav}-\omega_\textnormal{T}) + \delta) \cdot \tau$ that cancels the accumulated phase and applies a virtual detuning $\delta$.}

\black{The excited population of the qubit follows the Ramsey pattern of the exponential decay with the oscillations at the frequency $\delta$ (see Fig. \ref{sfig:cavity_T2_coherent_decay} (a)). Our data shows that the oscillations decay with the characteristic lifetime of 12\,$\mu$s. We present the hypotheses on what limits the cavity dephasing in the subsequent paragraphs.}


\black{Application of the same experiment for the large qubit-cavity detuning ($>1\,$GHz) is challenging due to the sensitivity of the swap operation to small deviations of the qubit frequency during the high-amplitude flux step pulse. This is caused by the distortion in the microwave line that includes a bias tee. As an alternative, we create a coherent state and perform the characteristic function tomography~\cite{campagne2020quantum, pan2023protecting} to fully reconstruct the evolution of the state in time. We solve the master equation for a given coherent state numerically for a range of the cavity dephasing rates and compare the measured characteristic functions with those of the simulated states. We use a gradient descent algorithm to find a value of the dephasing rate that yields the most similar characteristic functions of the simulated and measured states (see Fig. \ref{sfig:cavity_T2_coherent_decay} (b)). We bias the qubit in the lower sweetspot, offering the highest $T_2^\textnormal{echo}$, and apply a flux step pulse during the coherent state evolution to probe cavity $T_2$ at a range of detunings. Using this method we found that cavity $T_2$ with the qubit in the lower sweetspot ($\Delta$=1.3 GHz, $T_1 = 350\mu$s), is $\sim$ 230 $\mu$s.}


\black{We attribute those dephasing values to the poor thermalisation in this specific cooldown when the measurements were taken. Indeed, although we observed the qubit thermal population of $\sim$ 1\% in the cooldown when all the paper data was taken, in the more recent cooldown when we measured the cavity $T_2$s we found the thermal population to be in the range of 5\% to 14\%. Although we cannot concretely validate the source of this elevated effective temperature, our best educated guess is that the chip is poorly thermalised to the mixing chamber plate of the dilution refrigerator due to 9-months gap between the cooldowns where the device was not in a well-controlled environment.} 

\black{With this in mind, we will use a naive model to see if this hypothesis could explain the observed dephasing time. In general, the contribution to the cavity dephasing rate from coupling to the hot modes is given by the following equation~\cite{rigetti2012superconducting, reagor2016quantum}:}


\black{
\begin{equation}\label{eq:cav_dephasing}
        \Gamma_{\phi,c}^\textnormal{th} = \frac{\gamma_1}{2}\textnormal{Re}\Big[\sqrt{\big(1 + \frac{2i \chi}{\gamma_1} \big)^2 + \big(\frac{8i \chi n_\textnormal{th}}{\gamma_1} \big)} -1 \Big],        
\end{equation}}
where $\Gamma_{\phi,c}$ is the cavity dephasing rate caused by the thermal population in the qubit $n_\textnormal{th}$, $\chi$ is the dispersive coupling between the modes, and $\gamma_1$ is the energy decay rate in the qubit ($\gamma_1 = 1/T_1$).


\black{The total measured cavity $T_2$ decay is given by $1/T_{2,c} = 1/(2T_{1,c})+1/T_{\phi,c}$. We express the total dephasing rate in the cavity as }

\black{
\begin{equation}\label{eq:env_dephasing}
        \Gamma_{\phi,c} = \frac{1}{T_{2,c}} - \frac{1}{2T_{1,c}} = \Gamma_{\phi,c}^\textnormal{th} + \Gamma_{\phi,c}^\textnormal{env},
\end{equation}}
where $\Gamma_{\phi,c}^\textnormal{env}$ descibes all the other contributions to the dephasing excluding the qubit thermal population.

\black{We estimate the contribution to the cavity dephasing rate from the hot qubit in strong and weak dispersive regimes (applied flux $\Phi = 0.24\, \Phi_0$ and $\Phi = 0.50\, \Phi_0$, respectively). The measured system parameters in these bias points are listed in the Tables \ref{tab_supple:2026} and \ref{tab_supple:2026_cav}. We find $1/\Gamma_{\phi,c}(\Phi = 0.24\, \Phi_0) \approx $ 28 $\mu$s and $1/\Gamma_{\phi,c}(\Phi = 0.50\, \Phi_0) \approx $ 2.2 ms. The limit on the pure dephasing time $T_{\phi,c}^\textnormal{env} = 1/ \Gamma_{\phi,c}^\textnormal{env}$ due to environment is $\sim$ 26 $\mu$s in the strong and $\sim$ 320 $\mu$s in the weak dispersive regimes. This indicates that although poor thermalisation of the qubit can be a significant source of cavity dephasing, there are most likely other sources of dephasing as well. }
    

\black{At this point, we do not have a full model to accurately account for the source of this additional dephasing. One possible culprit is the environmental dephasing to the fabrication imperfections, in particular to two-level systems (TLS). We observed multiple two-level systems in the qubit frequency range, as evidenced directly by the avoided crossings at the frequencies where we don't expect any system eigenmodes from the electromagnetic simulations. An example of these avoided crossings is shown in Fig.~\ref{sfig:cavity_T2_coherent_decay} (c). Given the presence of these unstable TLS at frequencies close to the cavity and transmon, we suspect that they also have non-negligible coupling to the cavity mode. However, at this point, we do not have sufficient evidence to fully quantify this process and thus, this can only be viewed as an educated guess. Further investigations are being carried out in an updated device to uncover more insights related to the source of cavity dephasing such flux-integrated 3D bosonic cQED systems. }


\section{Calibration of transitions}

The Jaynes-Cummings ladder has a rich spectrum of transitions. We calibrate the frequency and amplitude of each transition by preparing the system in one of the adjacent states and then performing spectroscopy and Rabi oscillation experiments.
\begin{table}[ht]
\centering
\begin{tabular}{|c|c|c|c|}
\hline
Transition & $\omega_s/2\pi$ (GHz) & $\Omega_s/2\pi$ (MHz) & $\sigma$ (ns)\\
\hline \hline
$|0\rangle \leftrightarrow |1+\rangle$ & $6.8809$ & $29.0$ & $68$ \\
$|0\rangle \leftrightarrow |1-\rangle$ & $6.8566$ & $25.5$ & $52$ \\
$|1-\rangle \leftrightarrow |2+\rangle$ & $6.8987$ & $22.3$ & $44$ \\
$|1+\rangle \leftrightarrow |2-\rangle$ & $6.8402$ & $18.9$ & $48$ \\
$|2-\rangle \leftrightarrow |3+\rangle$ & $6.9074$ & $24.1$ & $44$ \\
$|2+\rangle \leftrightarrow |3-\rangle$ & $6.8315$ & $19.4$ & $32$ \\
$|3-\rangle \leftrightarrow |4+\rangle$ & $6.9142$ & $25.9$ & $44$ \\
$|3+\rangle \leftrightarrow |4-\rangle$ & $6.8246$ & $22.0$ & $24$ \\
$|4-\rangle \leftrightarrow |5+\rangle$ & $6.9201$ & $25.5$ & $32$ \\
$|5+\rangle \leftrightarrow |6-\rangle$ & $6.8143$ & $22.7$ & --- \\
$|6-\rangle \leftrightarrow |7+\rangle$ & $6.9291$ & $24.5$ & --- \\
\bottomrule
\addlinespace
\end{tabular}
\caption[Transition calibration parameters]{\textbf{Calibration of the parameters of the transitions drives.} The table shows the frequency of each relevant transition and its respective maximum Rabi frequency $\Omega_s$. $\Omega_s$ is defined as the coefficient of the $\frac{\Omega_s}{2}|\left(N+1\right)\mp\rangle\langle N\pm|$ matrix element when activated through the transmon drive. The parameter $\sigma$ is the standard deviation of the Gaussian $\pi$ pulse calibrated for each transition, with total duration of $4\sigma$. The length of the $|5+\rangle \leftrightarrow |6-\rangle$ and $|6-\rangle \leftrightarrow |7+\rangle$ transitions were not optimised.}
\label{tab:cal_sidebands}
\end{table}
For example, to calibrate the $|3-\rangle\leftrightarrow|4+\rangle$ transition the system is initialised in $|3-\rangle$ with the cavity and transmon on resonance. The transmon is then driven by a pulse with variable frequency and subsequently adiabatically detuned from the cavity. Finally, the transmon, now dispersively coupled to the cavity, is used to measure the parity using a standard Ramsey sequence. A change in parity indicates that drive has excited one of the four allowed transitions to states $|2\pm\rangle$ or $|4\pm\rangle$, which appear as four well-resolved features in the spectroscopy diagram. The desired $|3-\rangle\leftrightarrow|4+\rangle$ transition is identified as the one with highest frequency, in accordance with the theory.

After the transition frequency is identified, the amplitudes of excitation pulses are calibrated with power Rabi experiments. A gaussian pulse with standard deviation $\sigma$ and a total duration of $4\sigma$ is applied to the transmon line with variable amplitude. The amplitude corresponding to a $\pi$ pulse is the point in which the parity of the system is changed from odd to even. This experiment allows us to calculate the Rabi frequency $\Omega_s$ of the effective drive $\frac{\Omega_s}{2}|4+\rangle\langle3{-}|$. The corresponding frequencies and amplitudes of each JC transition are calibrated using this method and summarised in Tab.~\ref{tab:cal_sidebands}.

We use this method to calibrate the gaussian excitation pulses used for cavity state preparation in the main text. The $\sigma$ of each pulse was swept while tracking the parity of the state after a $2\pi$ rotation. The ideal $\sigma$ is such that the parity of the final state is as close as possible to the parity of the initial state. This corresponds to a compromise between the leakage and decoherence errors that affect short and long pulses, respectively.

\section{Decoherence model for simulation}

\subsection{Jump operators}

\black{For simulations, we take into account four error channels: transmon energy decay, transmon dephasing, cavity decay, and cavity dephasing.
The effect of the error channels is captured by the Lindblad master equation
\begin{equation}
    \dot \rho = -\frac{i}{\hbar}[H,\rho]+\sum_n L_n \rho L_n^{\dagger} -\frac{1}{2} (L_n^{\dagger}L_n \rho + \rho L_n^{\dagger}L_n),
    \label{EQ_Linblad}
\end{equation}
where $\{L_n\}$ are the jump operators corresponding to the error channels.
}

\black{
The jump operators for the transmon energy decay, transmon dephasing, cavity decay, and cavity dephasing are given by 
\begin{eqnarray}
    L_1&=& \sqrt{\frac{1}{T_1}} |g\rangle \langle e| \nonumber \\
    L_2&=& \sqrt{\frac{2}{T_{\phi}}} |e\rangle \langle e| \nonumber \\
    L_3&=& \sqrt{\frac{1}{T_{1,\text{cav}}}} a\nonumber \\
    L_4&=& \sqrt{\frac{2}{T_{\phi,\text{cav}}}} a^{\dagger}a, 
\end{eqnarray}
respectively.
}

\subsection{Transmon state transformation}

\black{For a transmon state undergoing energy decay alone for a time $T$, one can solve Eq.~(\ref{EQ_Linblad}) with a jump operator $L_1$. 
It is straight forward to obtain the following transformation
\begin{eqnarray}
    |g\rangle \langle g|&\rightarrow&|g\rangle \langle g|, \nonumber \\
    |e\rangle \langle e|&\rightarrow&|e\rangle \langle e|e^{-\frac{T}{T_1}}+|g\rangle \langle g|(1-e^{-\frac{T}{T_1}}), \nonumber \\
    |g\rangle \langle e|&\rightarrow&|g\rangle \langle e|e^{-\frac{T}{2T_1}}.\label{EQ_QT1}
\end{eqnarray}
}

\black{
For the case of a transmon undergoing pure dephasing alone, one can solve Eq.~(\ref{EQ_Linblad}) with a jump operator $L_2$ and obtain the following transformation
\begin{eqnarray}
    |g\rangle \langle e|&\rightarrow&|g\rangle \langle e|e^{-\frac{T}{T_{\phi}}},\label{EQ_QTphi}
\end{eqnarray}
where the diagonal elements are constant.
}

\subsection{Jump operators for dressed states}
\black{We analyse how the bare transmon and cavity jump operators manifest in the resonant JC picture by a simple change of basis to the JC dressed states. Having restricted the expression to the 1-excitation manifold for simplicity, we found that the effective operators on resonance can be written as} 

\black{
\begin{align}
        c_1 &= \sqrt{\frac{\gamma_1}{2}} \ket{0} \big[ (\bra{1+}) - (\bra{1-}) \big],\\
        c_2 &= \sqrt{\frac{\kappa^q_\phi}{2}} \sigma_x^{\text{dressed}},\\
        c_3 &= \sqrt{\frac{\kappa}{2}} \ket{0} \big[ (\bra{1+}) + (\bra{1-}) \big],\\
        c_4 &= \sqrt{\frac{\kappa^c_\phi}{2}}\sigma_x^{\text{dressed}},
\end{align}}

\black{where $\sigma_x^{\text{dressed}} =\big[ (\ket{1+}\bra{1-}) + (\ket{1-}\bra{1+})\big]$, and $\gamma_1$ and $\kappa$ are qubit and cavity energy loss rates, while $\kappa^q_\phi$ and $\kappa^c_\phi$ are qubit and cavity dephasing rates, respectively.}

\section{Comparison with other control methods}

Here, we compare the resonant control that we implemented to other well-known control techniques in the dispersive regime such as selective number-dependent arbitrary phase (SNAP) gate~\cite{krastanov_universal_2015,heeres_cavity_2015}, gradient ascent pulse engineering (GRAPE)~\cite{heeres_implementing_2017}, and echoed conditional displacement (ECD)~\cite{eickbusch_fast_2022}. 
We will make the comparison in two folds: in terms of the gate $T$ and effective lifetime $T_{\text{eff}}$ of the system, the summary of which is presented in Table~\ref{tab:t_and_Teff}.
Note that to make this comparison, we assume the following: 
\begin{enumerate}[label=(\roman*)]
    \item the physical system consists of a transmon and an oscillator;
    \item the task is to prepare an arbitrary $D$-dimensional pure state: $\sum_{n=0}^M c_n |n\rangle$, where $M=D-1$;
    \item multitone drives or numerical optimization of pulses are allowed.
\end{enumerate}

\subsection{Gate time}
\black{Here, the comparison will be made at the fundamental level without considering decoherence. The latter will be discussed in the next section.
Let us begin by noting that the dispersive regime is a special case of the JC regime where the detuning is much larger than the coupling $g$ ($\Delta=\omega_q-\omega_c\gg g$). In this regime, the main coupling term reads $\hat H_{\text{int}}=-\chi \hat a^{\dagger}\hat a |e\rangle\langle e|$. Notably, the dispersive regime can be further subdivided into two classes based on the ratio between $\chi$ and the fastest decoherence rate in the system, $\textnormal{max}(\Gamma_\phi, \Gamma_1, \kappa)$, characterising ancilla dephasing $T_\phi = 1/\Gamma_\phi$ and relaxation $T_1=1/\Gamma_1$, as well as cavity lifetime $T_{1,\textnormal{cav}}=1/\kappa$. 
In the strong dispersive regime, the bare nonlinearity $\chi$ is considerably larger than $\textnormal{max}(\Gamma_\phi, \Gamma_1, \kappa)$, whereas in the weak dispersive regime $\chi \lesssim \textnormal{max}(\Gamma_\phi, \Gamma_1, \kappa)$. 
}

\black{SNAP and GRAPE operate in the strong dispersive regime, where typically $\chi/2\pi \sim 0.5-2\:\textnormal{MHz}$. 
For SNAP, by driving the qubit at a certain frequency, an arbitrary phase can be imparted on the cavity Fock state $|n\rangle$ selectively, which essentially sets the gate time as $T\sim 1/\Omega$, where $\Omega$ is the strength of the qubit drive~\cite{krastanov_universal_2015}. 
For generating an arbitrary state within Fock state $|n\rangle $, it would involve applying the SNAP gate $n$ times, which gives $T\sim n/\Omega$ as the total gate time. 
However, with multi-tone qubit drives, the operation imparting arbitrary phases for all relevant Fock states can be applied simultaneously, making the gate time faster: $T\sim 1/\Omega$, which is independent of $n$. 
And it is important to point out that for SNAP, selectivity demands $\Omega \ll \chi$, and therefore the gate time scales as $T\sim 1/\Omega \gg 1/\chi$. 
}

\black{Another technique, GRAPE, does not require the strict selectivity of SNAP. 
Also, the qubit and cavity drives can be simultaneously applied while the qubit and cavity are dispersively coupled. 
This way, GRAPE is more general compared to SNAP, i.e., the solution space for GRAPE contains that of SNAP. 
This makes the gate time for GRAPE often faster than multi-tone SNAP, which is set by the dispersive coupling strength $T\sim 1/\chi$~\cite{ma2021quantum}.
For this reason, we will omit SNAP hereafter.
}

\black{Next, ECD is a control technique operating in the weak dispersive regime with lower $\chi$. This regime, where $\chi$ is made small intentionally ($\approx 10-100\,$kHz), is preferred when higher-order nonlinearities in the cavity mode is a key limiting factor for the intended task. 
For a single ECD gate, the time scales as $T_1\sim |\beta|/\chi\alpha$~\cite{eickbusch_fast_2022}, where $|\beta|$ is the \emph{size} of the ECD gate and $\alpha$ is the cavity displacement applied during the gate.
For the task of preparing the arbitrary state, an approximation can be made on the average photon number: in order to prepare an arbitrary state with mean photon number $M/2$, one requires the size of the ECD to be $\sim|\beta/2|^2$, giving $|\beta|\sim \sqrt{M}$ scaling.
Furthermore, a pure state of dimension $D=M+1$ has $2M$ real parameters.
A single unit consisting of an ECD and a qubit rotation gives $4$ real parameters~\cite{eickbusch_fast_2022}, which means that for our task one requires $\sim M/2$ applications of the units.
Therefore, the total gate time scales as $T\sim M^{3/2}/(\chi\alpha)$.
Compared to GRAPE, the gate time for ECD suffers from the $M$ dependence and lower $\chi$ value. However, by leveraging a large amplitude $\alpha$, the ECD can achieve comparable gate time to GRAPE (in the strong dispersive regime) and faster if the comparison is made in the same regime of low $\chi$~\cite{eickbusch_fast_2022}. The bound on the amplitude $\alpha$ can be associated with the condition to operate with an average photon number $\bar n$ ($\approx \alpha^2$) much smaller than the critical photon number allowed in the system, which gives $\chi \alpha \ll \sqrt{\chi K/6}$~\cite{eickbusch_fast_2022}, where $K$ is the transmon anharmonicity, such that the gate time now follows $T\gg M^{3/2} \sqrt{6/(\chi K)}$.
}

\begin{figure}[ht]
\centering
\includegraphics{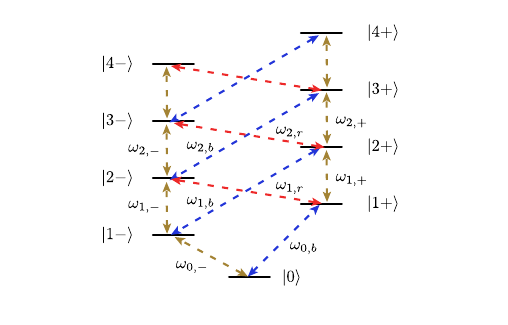}
\caption{\black{JC ladders and their transition frequencies: $\omega_{N,r}=-g(\sqrt{N+1}+\sqrt{N})$, $\omega_{N,b}=g(\sqrt{N+1}+\sqrt{N})$, and $\omega_{N,\pm}=\pm g(\sqrt{N+1}-\sqrt{N})$. For multitone transmon drive with numerical optimization, we simultaneously drive transition frequencies $\{\omega_{0,-},\omega_{N,r},\omega_{N,b}\}$, which is sufficient to reach all the JC states.}}
\label{fig:JC_ladder}
\end{figure}

\black{For the resonant control technique, we take the case where we use multitone transmon drive addressing all relevant JC transitions (illustrated in Fig.~\ref{fig:JC_ladder}) simultaneously~\cite{strauch_resonant_2012}.
}


\black{To address a particular transition, e.g., $|M-\rangle \leftrightarrow|(M+1)+\rangle$, the duration follows $T_M\sim1/\Omega$ with $\Omega$ being the transmon drive amplitude. 
The drive amplitude cannot be too high (narrow in time or broad in the frequency domain) as this may excite unwanted transitions. As the energy separation between adjacent sidebands scales as $g/\sqrt{M}$, a selectivity condition requires $\Omega\sim g/\sqrt{M}$.
Therefore $T_M\sim \sqrt{M}/g$.
For the case of simultaneously driving multiple relevant transitions with multitone drives to create the arbitrary state, the gate time then follows $T\sim \max_M T_M=\sqrt{M}/g$.}

\black{In the dispersive regime, the characteristic frequency separation is $2\chi = 2 g^2 / (2\pi \Delta)$ (considering the two-level approximation of the ancilla qubit), where $\Delta$ is the detuning between the qubit and the cavity in the units of Hz.
The dispersive approximation requires the number of photons in the oscillator $M$ to be much smaller than the critical photon number, $M\ll N_{\textnormal{crit}} = \big((2\pi \Delta) / 2g \big)^2$ \cite{blais_circuit_2021}. Hence, $T_M = \sqrt{M}/g \ll \frac{2\pi \Delta}{2g^2} = 1/2\chi$.
Therefore, within the range of photon numbers attainable in the dispersive regime, characteristic gate times offered by the resonant control protocol are faster than those for the dispersive techniques.
}

\black{To proceed with the comparison, we will take typical values of hardware parameters and evaluate the expressions in the second column of Table~\ref{tab:t_and_Teff}.
Note that the expressions are simply scaling; the real gate time should come with a prefactor (here $T^*$ is computed with a prefactor of $2\pi$ for simplicity).
Strong dispersive regime where GRAPE is implemented typically operates with $\chi/2\pi \sim 0.5-2$~MHz, making $T^* \approx$ 100s-1000s~ns.
To assess the case of ECD, we note that typical coupling strength in the weak dispersive regime is $\chi/2\pi\sim 10-100$~kHz and transmon anharmonicity is $K/2\pi\sim 200-300$~MHz. 
This approximates the \emph{lower bound} for the gate time $T^*\approx M^{3/2}\times$ 100s-1000s~ns .
Finally, for the resonant regime, the the coupling strength ranges from $g/2\pi\sim 10$~MHz to $500$~MHz~\cite{kurilovich2025high}, \black{which makes $T^*\approx M^{1/2}\times$ 10s-100s~ns}.
From this initial assessment, we conclude that for a device with a given $g$, the resonant control method \emph{in principle} provides the fastest gate time compared to other methods in the dispersive regime.
}

\subsection{Effective lifetime}\label{S_effective_lifetime}

\black{Here, we will assume that the main source of decoherence consists of transmon energy decay and dephasing, which are respectively characterised by $T_1$ and $T_{\phi}$. 
Also, we consider the gate time $T\ll T_1,T_{\phi}$ such that the decoherence channels can be approximated to first order. 
We will present the analytical evidence for the approximation of the resulting effective lifetime of the system during the same task of preparing the arbitrary state of dimension $D=M+1$ and show that the expressions are comparable for the different methods (summarized in the fourth column of Table~\ref{tab:t_and_Teff}).
This gives the resonant control an intrinsic advantage, i.e., it has a lower ratio between the gate time and effective lifetime, which is directly related to infidelity. 
}

\subsubsection{Resonant control $T_{\text{eff}}$}

\black{For resonant control the preparation of an arbitrary state is achieved by simultaneously driving relevant transitions of the JC ladders (Fig.~\ref{fig:JC_ladder}). To approximate the lifetime of the system, let us assume that at some point in time, the system is in a superposition of all JC states
\begin{widetext}
\begin{eqnarray}
    |\psi\rangle&=&\alpha|M+\rangle + \beta|M-\rangle + \gamma|(M-1)+\rangle + \delta|(M-1)-\rangle + \cdots \nonumber \\
    &=&\frac{1}{\sqrt{2}}\big(|g\rangle(\alpha|M\rangle+\beta|M\rangle+\gamma|M-1\rangle+\delta|M-1\rangle+\cdots) +|e\rangle(\alpha|M\rangle-\beta|M\rangle+\gamma|M-1\rangle-\delta|M-1\rangle+\cdots)\big) \nonumber \\
    &=&\lambda_g |g\rangle |u_g\rangle + \lambda_e |e\rangle |u_e\rangle,\label{EQ_resonant_allJCstate}
\end{eqnarray}
\end{widetext}
where $M=D-1$, 
\begin{eqnarray}
    \lambda_g &=&\sqrt{ (1+2\text{Re}(\alpha \beta^*)+2\text{Re}(\gamma \delta^*)+\cdots)/2} \nonumber \\
    \lambda_e &=&\sqrt{ (1-2\text{Re}(\alpha \beta^*)-2\text{Re}(\gamma \delta^*)-\cdots)/2},
\end{eqnarray}  
with $|u_g\rangle$ and $|u_e\rangle$ being the normalised states of the cavity corresponding to the transmon in $|g\rangle$ and $|e\rangle$, respectively.
}

\black{
Under transmon energy decay, the transmon state transformation in Eq.~(\ref{EQ_QT1}) can be applied to the density matrix version of Eq.~(\ref{EQ_resonant_allJCstate}) to arrive at
\begin{eqnarray}
    \rho&=&|\lambda_g|^2 |g\rangle \langle g||u_g\rangle \langle u_g|+\lambda_g\lambda_e^* |g\rangle \langle e||u_g\rangle \langle u_e|e^{-\frac{T}{2T_1}}\nonumber \\
    &&+\lambda_g^*\lambda_e |e\rangle \langle g||u_e\rangle \langle u_g|e^{-\frac{T}{2T_1}}+|\lambda_e|^2 |e\rangle \langle e||u_e\rangle \langle u_e|e^{-\frac{T}{T_1}}\nonumber \\
    &&+|\lambda_e|^2 |g\rangle \langle g||u_e\rangle \langle u_e|(1-e^{-\frac{T}{T_1}}).\label{EQ_decayRC_dm}
\end{eqnarray}
The fidelity with state $|\psi\rangle$ then reads
\begin{eqnarray}
    F&=&\langle \psi|\rho|\psi\rangle \nonumber \\
    &\approx&1-\frac{\tilde p_e^2T}{T_1}-\frac{\tilde p_g\tilde p_eT}{T_1}+\frac{\tilde p_g\tilde p_e|\langle u_g|u_e\rangle|^2T}{T_1}\nonumber \\
    &\equiv&1-\epsilon_1
\end{eqnarray}
where we have used $\tilde p_g=|\lambda_g|^2$, $\tilde p_e=|\lambda_e|^2$, and the approximation is valid to first order.
The error $\epsilon_1$ is interpreted as infidelity, which is \textit{upperbounded} as  
\begin{equation}
    \epsilon_1 =\frac{\tilde p_e^2T}{T_1}+\frac{\tilde p_g\tilde p_eT}{T_1}-\frac{\tilde p_g\tilde p_e|\langle u_g|u_e\rangle|^2T}{T_1}\le \frac{\tilde p_eT}{T_1}.
\end{equation}
}

\black{
Similar analysis can be done considering transmon pure dephasing only, in which the fidelity can be written as
\begin{eqnarray}
    F&=&\langle \psi|\rho|\psi\rangle \nonumber \\
    &\approx&1-\frac{2\tilde p_g\tilde p_eT}{T_{\phi}}\nonumber \\
    &\equiv&1-\epsilon_2.
\end{eqnarray}
}

\black{To first order the errors add up $\epsilon=\epsilon_1+\epsilon_2$ (total infidelity) in the presence of both decoherence channels and we can write the final expression as
\begin{eqnarray}
    F&=&1-\epsilon\nonumber \\
    F&\gtrsim&1- \frac{\tilde p_eT}{T_1}-\frac{2\tilde p_g\tilde p_eT}{T_{\phi}}\nonumber \\
    &\equiv&1-\frac{T}{T_{\text{eff}}},
\end{eqnarray}
from which the effective lifetime of the system is written as $T_{\text{eff}}=(2\tilde p_g\tilde p_e/T_{\phi}+\tilde p_e/T_1)^{-1}$.
We note that, as the transitions $|M+\rangle \leftrightarrow |M-\rangle$, $|(M-1)+\rangle \leftrightarrow |(M-1)-\rangle$, $\cdots$ are never driven, their simultaneous populations are nearly zero most of the time.
In other words, $\text{Re}(\alpha \beta^*)\approx 0$, $\text{Re}(\gamma \delta^*)\approx 0$, $\cdots$, such that $\tilde p_e\approx \tilde p_g\approx 0.5$ and
\begin{equation}
    T_{\text{eff}}=(1/(2T_{\phi})+1/(2T_1))^{-1}.
\end{equation}
}

\subsubsection{GRAPE $T_{\text{eff}}$}

\black{For the case of GRAPE, the protocol is hard to write analytically as it allows arbitrary smooth time-dependent drives for both the transmon and oscillator. In order to approximate the effective lifetime, we start with an arbitrary state 
\begin{equation}
    |\psi \rangle = \alpha |g\rangle |\psi_g\rangle + \beta |e\rangle |\psi_e\rangle,
\end{equation}
where $|\psi_g\rangle$ and $|\psi_e\rangle$ are oscillator states corresponding to the transmon state in $|g\rangle$ and $|e\rangle$, respectively.
Under transmon energy decay and dephasing, the state will decay similar to the process written in Eq.~(\ref{EQ_decayRC_dm}).
The rest of the analysis follows similarly, after which the fidelity to the initial state is expressed as
\begin{equation}
    F\gtrsim 1-\frac{2p_gp_eT}{T_{\phi}}-\frac{p_e T}{T_1}, 
\end{equation}
where $p_g=|\alpha|^2$ and $p_e=|\beta|^2$, resulting in 
\begin{equation}
    T_{\text{eff}}=(2p_gp_e/T_{\phi}+p_e/T_1)^{-1}.
\end{equation}
}

\subsubsection{ECD $T_{\text{eff}}$}

\black{Lastly for ECD, the preparation of an arbitrary state requires multiple rounds of an ECD and a qubit rotation~\cite{eickbusch_fast_2022}. We will only scrutinize decoherence during the ECD operations as the qubit operations are normally much faster. Also, as the exact sequence varies depending on a particular task and numerical optimization, we will approximate the effective lifetime during a representative operation: single ECD applied to an initial state $(|g\rangle +|e\rangle)|0\rangle/\sqrt{2}$.
Similar with previous analysis, the state will undergo transmon decoherence, during the entire ECD sequence that involves phase-space trajectories for the cavity state conditional on the transmon state~\cite{eickbusch_fast_2022}.
The ECD gate with total duration $T$ (for the sake of notation) is formally written as $ECD(\beta)=|e\rangle \langle g|D(\beta/2)+|g\rangle \langle e|D(-\beta/2)$. 
Following the applied sequence, however, it is more meaningful to break it down to 3 operations: (i) conditional gate $G_1=|g\rangle \langle g|\tilde D(\beta/4)+|e\rangle \langle e|\tilde D(-\beta/4)$ for $T/2$; (ii) fast qubit flip $|g\rangle \leftrightarrow |e\rangle$; and another conditional gate $G_2=|g\rangle \langle g|\tilde{\tilde D}(-\beta/4)+|e\rangle \langle e|\tilde{\tilde D}(\beta/4)$ for $T/2$. The single and double tilde notations are used as a reminder that these are not displacement operators; they are phase-space trajectories that would effectively displace cavity state the same way as normal displacement operators in the ideal case.
}

\black{
Without decoherence, the state of the system after ECD is $|\psi\rangle=(|g\rangle |-\beta/2\rangle+|e\rangle |\beta/2\rangle)/\sqrt{2}$.
Under qubit energy decay alone, the state after $G_1$ and after $G_2$ are written as 
\begin{widetext}
\begin{eqnarray}
    \rho_1&=&\frac{1}{2}\big( |g\rangle \langle g||\frac{\beta}{4}\rangle \langle \frac{\beta}{4}| + |g\rangle \langle e||\frac{\beta}{4}\rangle \langle -\frac{\beta}{4}|e^{-\frac{T}{4T_1}} + |e\rangle \langle g||-\frac{\beta}{4}\rangle \langle \frac{\beta}{4}|e^{-\frac{T}{4T_1}}+ |e\rangle \langle e||-\frac{\beta}{4}\rangle \langle -\frac{\beta}{4}|e^{-\frac{T}{2T_1}}\nonumber \\
    &&+ |g\rangle \langle g||-\frac{\beta}{4}\rangle \langle -\frac{\beta}{4}|(1-e^{-\frac{T}{2T_1}}) \big) \nonumber \\
    \rho_2&=&\frac{1}{2}\big( |g\rangle \langle g||-\frac{\beta}{2}\rangle \langle -\frac{\beta}{2}|e^{-\frac{T}{2T_1}} + |g\rangle \langle e||-\frac{\beta}{2}\rangle \langle \frac{\beta}{2}|e^{-\frac{T}{2T_1}} + |e\rangle \langle g||\frac{\beta}{2}\rangle \langle -\frac{\beta}{2}|e^{-\frac{T}{2T_1}}\nonumber \\
    &&+ (|e\rangle \langle e||\frac{\beta}{2}\rangle \langle \frac{\beta}{2}|+|e\rangle \langle e||0\rangle \langle 0|(1-e^{-\frac{T}{2T_1}}))e^{-\frac{T}{2T_1}}\big) \nonumber \\
    &&+ (|g\rangle \langle g||\frac{\beta}{2}\rangle \langle \frac{\beta}{2}|+|g\rangle \langle g||0\rangle \langle 0|(1-e^{-\frac{T}{2T_1}}))(1-e^{-\frac{T}{2T_1}})\big),
\end{eqnarray}
\end{widetext}
respectively.
}

\black{The fidelity to the ideal target state then reads 
\begin{eqnarray}
    F&=&\langle\psi|\rho_2|\psi\rangle \nonumber \\
    &\approx& 1-\frac{T}{2T_1},
\end{eqnarray} 
which gives $\epsilon_1=\frac{T}{2T_1}$.
Similar analysis considering tramson dephasing alone yield $\epsilon_2=\frac{T}{2T_{\phi}}$
Consequently, the effective lifetime is given by 
\begin{equation}
    T_{\text{eff}}=(1/(2T_{\phi})+1/(2T_1))^{-1}.
\end{equation}
}

\black{In conclusion, with transmon decoherence as the main detrimental factor, the expressions for the effective lifetime are comparable for resonant control, GRAPE, and ECD, especially for the case where the transmon excited state is used half the time ($p_e=p_g=0.5$) in GRAPE.
Therefore, the resonant control retains the intrinsic advantage by having the lowest gate time to lifetime ratio $T/T_{\text{eff}}$, a figure of merit directly translates to infidelity. 
}

\black{
We note that the effective lifetime can be extended by implementation of error-corrected gates that are fault-tolerant to ancilla decoherence \cite{reinhold2020error, tsunoda2023error, roy2026error}. The implementation of such gates within the resonant control protocol is an interesting direction for future investigation.
}

\subsection{Other factors}

\black{
We \black{expect that} the reported fidelities $\approx0.68-0.93$ in this work are not fundamentally limited by the resonant control technique.
The main limitation here is the qubit decoherence
and the relatively low coupling $g/2\pi\approx 12$~MHz, both of which can be improved with better design and fabrication.
Based on our analysis above, the gate time for one JC transition for this fixed $g$ is approximately a few $100$s~ns. Furthermore, during the operation of $500-800$~ns, our analytical model approximates the upper bound on infidelity as $T(1/(2T_{\phi})+1/(2T_1))\approx 14-22\,\%$. 
We also added an additional model to estimate the upper bound of the error during the adiabatic detuning (detailed Sec~\ref{sec_adiabaticdetuning}), which may add up to $6\%$ infidelity. 
In summary, the lower bound on the expected fidelity of $500-800$~ns gate with adiabatic detuning is $0.72-0.80$, which is in line with the reported experimental values.
For improved qubit coherence of $T_1=100\,\mu$s and $T_{\phi}=50\,\mu$s, our analytical model predicts an upper bound infidelity of $2\%$ for the cases described above.
}

\black{
Furthermore, the benefit of resonant control can be more prominent for higher coupling $g$. 
Typical values of $g/2\pi$ currently ranges from $10$ to $500$~MHz~\cite{kurilovich2025high} ($12$~MHz in this work), where the maximum value would correspond to gate time $\sim10$s~ns.
Increasing $g$ further is an ongoing research effort, with the potential to reach the regime of ultra strong or even deep strong coupling.
It is well-known that the transition from the resonant JC to ultra strong, where the effects of the counterrotating terms in the interaction Hamiltonian become observable, starts at $g/\omega_c \approx 0.1$~\cite{forn2019ultrastrong}.
Given that the oscillator's frequency is $\approx 6.9$~GHz, this would give the upper bound $g_{\text{max}}/2\pi\approx 690$~MHz (gate time $T\sim 1$s~ns), in which the resonant control regime can still be valid. From the current coupling value ($12$~MHz), there is still a lot of room before the resonant control breaks.
}

\black{However, when the coupling strength exceeds the transmon anharmonicity, the two-level approximation may break down, leading to spectral crowding arising from transitions involving higher excited transmon states. This motivates exploring alternative ancilla qubits with substantially larger anharmonicities, such as flux qubits \cite{yan2016flux} and fluxonium \cite{nguyen2019high,atanasova_insitu_2025}}.

\subsection{Comparison using numerical simulations}

\black{We use numerical simulations to demonstrate the validity of our theoretical analysis in the sections above. First, we compare all-resonant control protocol to GRAPE~\cite{heeres_implementing_2017}, choosing the task of preparing the state $\tfrac{1}{\sqrt{2}}\left(|0\rangle + |3\rangle\right)$, which was experimentally demonstrated in this work (main text, Fig.~3 (a)).}

\black{For the all-resonant control, we simulate the system with an anharmonic 4-level ancillary transmon coupled to the bosonic cavity with the Jaynes-Cummings Hamiltonian \eqref{eq:JC_Hamiltonian_lab_frame}. For GRAPE, we use the dispersive Hamiltonian with the cross-Kerr between the transmon and cavity $\chi/2\pi$ = $-$1 MHz, typically used for GRAPE implementations. To find the higher-order correction to the Hamiltonian, we numerically diagonalise Jaynes-Cummings Hamiltonian \eqref{eq:JC_Hamiltonian_lab_frame}, find the transmon frequency that gives $\chi/2\pi$ = $-$1 MHz, and find the second-order dispersive shift $\chi'$ and cavity self-Kerr from the Hamiltonian eigenenergies. The static Hamiltonian for the case of GRAPE (in the frame rotating with the qubit and cavity) is}

\black{
\begin{equation}
    \hat{H}_{\textnormal{disp}} = \frac{K}{2} \hat{b}^\dagger \hat{b}^\dagger \hat{b} \hat{b} + \chi \hat{a}^\dagger \hat{a} \vert{}e\rangle\langle e\vert{} + \frac{\mathcal{K}}{2} \hat{a}^\dagger \hat{a}^\dagger \hat{a} \hat{a} + \frac{\chi'}{2} \hat{a}^\dagger \hat{a}^\dagger \hat{a} \hat{a} \vert{}e\rangle\langle e\vert{},
\end{equation}}

\black{where $\hat{a}$ and $\hat{b}$ are cavity and transmon annihilation operators, respectively, $K$ is the transmon anharmonicity, $\mathcal{K}$ is the cavity self-Kerr, and $\chi'$ is the second-order dispersive shift.} 

\black{We find that for the parameters from the Table \ref{tab_supple:device_parameters}, $\chi/2\pi$ = $-$1 MHz is achieved at $\Delta$ = 161 MHz, with $\mathcal{K}/2\pi$ = $-$3.9 kHz and $\chi'/2\pi$ = 7.5 kHz.}

\black{We optimised the time-dependent cavity and transmon drives using the GRAPE algorithm with a target state fidelity of $0.99$ (the optimisation is conducted for the system with no decoherence). We take the duration as the minimum time for which the optimisation still converges.}

\black{We apply the optimised GRAPE pulses on the lossy system with the range of qubit lifetimes $T_1 \in [$10, 15$]$ $\mu$s and $T_2 \in [$1.3, 2.8$]$ $\mu$s, while cavity $T_1$ is fixed to 500 $\mu$s. It yields fidelities in the range 0.779-0.874 for the 1000~ns state preparation, while all-resonant control with sequential drive reaches fidelities 0.814-0.890 in 640~ns.
We also found that all-resonant fidelities can be improved to between 0.832 and 0.913 within the same state preparation time by tuning the durations and amplitudes of the drive pulses used to climb the JC ladder, effectively optimizing the sequence to mitigate coherent errors.}

\black{For enhanced qubit lifetimes $T_1 = 100\,\mu\text{s}$, $T_\phi = 50\,\mu\text{s}$, the optimised all-resonant pulse sequence reaches fidelity of 0.990, while GRAPE yields 0.972.}

\black{To compare our protocol to ECD, we simulate the preparation of the Fock state $\ket{5}$, as it was realised experimentally in both our work (main text, Fig.~3 (c)) and Ref.~\cite{eickbusch_fast_2022}. 
Using numerical simulations, we find that the fidelity of preparing the Fock state $|5\rangle$ in 800~ns with the multi-tone drive is between 0.70 and 0.83 for the previously considered range of qubit lifetimes. This is consistent with our experimentally measured fidelity of $0.75 \pm 0.02$.
Improving the transmon coherence to the same values as in Ref.~\cite{eickbusch_fast_2022} ($T_1\approx50\,\mu$s and $T_2\approx30\,\mu$s), with all other device parameters held constant would improve the state preparation fidelity to $0.97$. In Ref.~\cite{eickbusch_fast_2022}, the simulated state fidelity for Fock $|5\rangle$ is $0.94$ with duration close to $2\,\mu$s.}

\black{Thus, full numerical simulations validate our theoretical analysis, demonstrating that all-resonant control operates on faster timescales than protocols relying on the dispersive regime, while maintaining a similar effective coherence time.}

\subsection{Discussion}

An important point worth stressing is that we detune the qubit after the protocol such that the qubit is decoupled from the cavity, and therefore the cavity's lifetime is back to its original value.
Even though the resonant control protocol introduces the small overhead due to the time required to adiabatically detune the system from resonance, this overhead is constant (does not scale with the gate complexity) and reasonably short even on the timescale of the current system lifetimes ($\sim\textnormal{200}$ ns).

The resonant scheme achieves fast gate time without compromising the lifetime of the system. Therefore, we \black{expect} that the JC protocol offers the most favourable strategy for universal control of a bosonic mode.

The deep-strong (DS) and ultra-strong (US) coupling regimes of the JC Hamiltonian, providing access to significantly higher values of coupling strength $g$, can be explored to improve our method further. However, the rotating wave approximation of the quantum Rabi model~\cite{mezzacapo_digital_2014}, used in current analysis (see Eq.~(1) of the main text), will not be applicable anymore in those regimes. Therefore, new strategies are to be developed for universal control of JC Hamiltonian in DS and US regimes.

Other methods, such as tailored nonlinear Hamiltonians beyond JC dynamics, can also be explored. For instance, 
the cubic interactions~\cite{eriksson2024universal} have been demonstrated on a low-Q oscillator by directly terminating the planar cavity with a nonlinear element. The strong nonlinearity inherent to such bosonic mode allows operations on the timescale of tens of nanoseconds. However, such an element is not an effective quantum memory, as the nonlinearity cannot be dynamically decoupled from the cavity, leading to always-on high decoherence rate. Furthermore, this scheme can only operate in the bounded area of the phase space ($|\alpha|<1.5$), outside of which the control is undermined by the static Kerr nonlinearity. Therefore, at this stage such an approach still requires further investigation in order to be a general-purpose tool to achieve universal control of a bosonic mode.


\begin{table*}[ht]
    \centering
    \begin{tabular}{|c|c|c|c|}
        \hline
        \textbf{Techniques} & \textbf{Gate time ($T$)} & \textbf{Gate time* ($T^*$)} & \textbf{Effective lifetime ($T_{\text{eff}}$)}  \\
        \hline
        SNAP & $\gg\frac{1}{\chi}$& 100s-1000s ns &- \\
        \hline
        GRAPE & $\sim\frac{1}{\chi}$ & 100s-1000s ns&$\sim (\frac{p_e}{T_1}+\frac{2p_ep_g}{T_{\phi}})^{-1}$ \\
        \hline
        ECD & $\sim M^{3/2}\frac{1}{\chi \alpha}\gg M^{3/2} \sqrt{\frac{6}{\chi K}}$ &$M^{3/2}$ 100s-1000s ns&$\sim (\frac{1}{2T_1}+\frac{1}{2T_{\phi}})^{-1}$ \\
        \hline
        Resonant & \black{$\sim\frac{M^{1/2}}{g}$}&\black{$M^{1/2}$} 10s-100s ns &$\sim (\frac{1}{2T_1} +\frac{1}{2T_{\phi}})^{-1}$\\
        \hline
    \end{tabular}
    \caption{Comparison of gate time at the fundamental level (second column) and considering typical hardware parameters (third column) for preparing an arbitrary oscillator state of dimension $D=M+1$. We also compare effective lifetime of the system (fourth column) given that a qubit with lifetimes $T_1$ and $T_{\phi}$, respectively characterizing the energy decay and dephasing, is used in the process.}
    \label{tab:t_and_Teff}
    \end{table*}

\section{Preparation of arbitrary bosonic states}\label{sec:states_preparation}
\begin{figure}[h!]
    \centering
    \includegraphics[width=\linewidth]{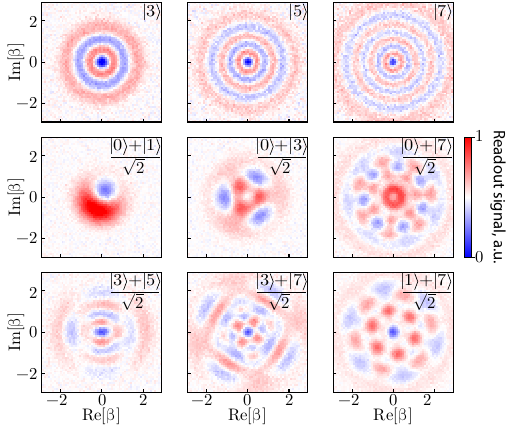}
    \caption[Qudit arbitrary state preparation]{\textbf{Qudit arbitrary state preparation.} Wigner plots showing diverse qudit states prepared by sequentially exciting sideband transitions in the Jaynes-Cummings ladder. }
    \label{sfig:qudit_preparation}
\end{figure}

We used the resonant control protocol to demonstrate the preparation of arbitrary states in the cavity with an analytical control sequence. The sequence consisted of activating the sidebands sequentially using variable-amplitude, Gaussian-shaped pulses with $4\sigma$ duration, with the parameters given in the Tab.~\ref{tab:cal_sidebands}. This method allows for straightforward preparation of any Fock state and their superpositions in the cavity, as shown in Fig.~\ref{sfig:qudit_preparation}.

Due to the lack of appropriate quantum amplifier at the specific moment of this experiment, we don't have single-shot readout of the transmon. Thus, the parity signal was calibrated by comparing it to a $1\textnormal{D}$ Wigner measurement of cavity vacuum $|0\rangle$, which is independently measured beforehand. 
\black{
We note that a 2\% error in the calibration scaling results in small errors ($<1$\%) in all the reported fidelities in the main text.} The Wigner plots closely resemble the target states in shape, though with a reduction in contrast. This indicates that the state preparation is mostly limited by the transmon decoherence.

\section{Arbitrary gate on logical states of binomial encoding}

As shown in Sec.~\ref{sec:states_preparation}, resonant control protocol enables arbitrary state preparation.
Here, we show that an arbitrary gate on the logical states of binomial encoding of the lowest order~\cite{michael_new_2016} can be achieved within the framework of the resonant control protocol. This follows as Givens rotations can be utilized to realise $\hat{R}_{x}^\textnormal{L}(\theta)$, $\hat{R}_{y}^\textnormal{L}(\theta)$, and $\hat{R}_{z}^\textnormal{L}(\theta)$ on the subspace $\big\{|\mydownarrow\rangle^\textnormal{L} = |2\rangle,  |\myuparrow\rangle^\textnormal{L} = (|0\rangle+|4\rangle)/\sqrt{2} \big\}$.
The rotations can be written explicitly in terms of Givens rotations as
\begin{subequations}\label{eq:binomial_rotation}
    \begin{equation}
    \hat{R}_{x}^\textnormal{L} (\theta) = \hat{\mathcal{R}}_{0,2}(\zeta_1) \hat{\mathcal{R}}_{2,4}(\zeta_2)\hat{\mathcal{R}}_{0,2}(\zeta_1),
    \end{equation}
    \begin{equation}
    \hat{R}_{y}^\textnormal{L} (\theta) = \hat{\mathcal{R}}^{-\pi/2}_{0,2}(\zeta_1) \hat{\mathcal{R}}^{\pi/2}_{2,4}(\zeta_2)\hat{\mathcal{R}}^{-\pi/2}_{0,2}(\zeta_1),
    \end{equation}
    \begin{equation}
        \hat{R}_{z}^\textnormal{L}(\theta) = \hat{\mathcal{R}}_{0,4}(-2\theta),
    \end{equation}
\end{subequations}
where the parameters $\zeta_1 = 2\arctan \big(\tan(\theta/4)/\sqrt{2} \big)$, $\zeta_2 = 4\arcsin \big(\sin(\theta/4)/\sqrt{2} \big)$, and the notation $\hat{\mathcal{R}}^{\gamma}_{n_A,n_B}(\alpha)$ means Givens rotation between the Fock states $|n_A\rangle$ and $|n_B\rangle$ by an angle $\alpha$ as described in the main text, where the pulse is phase-shifted by $\gamma$.



Thus, arbitrary control on the logical states of the binomial encoding is enabled via trivial combinations of the Givens rotations.

\section{Parametric modulation}

Although not required for our resonant control protocol, our device architecture is also capable of providing broadband parametric flux modulation of the SQUID. 

\begin{figure}[bh!]
\centering
\includegraphics{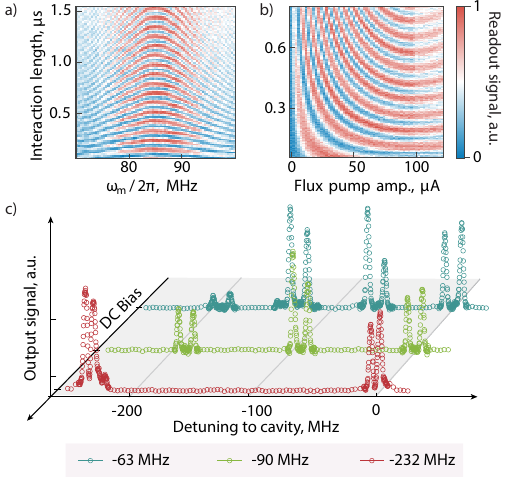}
\caption{\textbf{Protected flux-parametric modulation.} \textbf{(a)} Vacuum Rabi oscillations activated by a flux sideband. The transmon-cavity detuning is $\Delta=\textnormal{85 MHz}$. After being excited with a $\pi$ pulse, a flux pump with $\omega_m/2\pi \approx \Delta$ creates a sideband that activates the photon exchange between transmon and cavity for a variable interaction time. \textbf{(b)} Resonant vacuum Rabi oscillations as a function of the flux pump amplitude. The speed of the Rabi oscillations can be continuously tuned in the $g_\textnormal{eff}/2\pi = \textnormal{0} \div \textnormal{6.57 MHz}$ range. \textbf{(c)} Transmon  spectroscopy for the variable values of $\Delta$ under a flux pump $\varphi_e \propto \cos\big(2\pi \Delta t \big)$. The flux pump activates the parametric resonance between the transmon and the cavity, as seen from the frequency split that appears on every sideband due to the avoided crossing.}
\label{fig:sfig_avoided_crossings}
\end{figure}

We show this feature experimentally by using flux pumps to activate parametric resonances between the cavity and transmon with tunable strength. Sweeping the pump frequency $\omega_m$ around the transmon-cavity detuning of $\Delta\approx85\,\textnormal{MHz}$ for a variable duration results in the continuous exchange of the transmon excitation with the cavity, as shown by the chevron pattern in Fig.~\ref{fig:sfig_avoided_crossings}(a). The resonant oscillations have a period of $\pi/g_\textnormal{eff}$, where $g_\textnormal{eff}$ is the parametric coupling strength.

The parametric coupling strength $g_{\textnormal{eff}}$ can be tuned on demand by changing the flux pump amplitude. Fig.~\ref{fig:sfig_avoided_crossings}(b) shows that the resonant oscillations become faster with larger flux pump amplitude (expressed in units of current), corresponding to an increase in coupling strength. The pump amplitude $I_p$ is estimated from the room-temperature input current after going through $26\,\textnormal{dB}$ attenuation and treating the flux line as a short. Fitting the oscillations to a cosine function for each pump amplitude indicates a variable $0\leq g_\textnormal{eff}\leq 6.57\,\textnormal{MHz}$. However, increasing the pump amplitude beyond the maximum coupling point leads to lower $g_\textnormal{eff}$. This is consistent with the expected analytical trend $g_\textnormal{eff} = g\times J_1(I_p/\nu )$, where $g$ is the static coupling, $J_1$ is a Bessel function of the first kind and $\nu$ is proportionality factor~\cite{didier2018analytical}. 

The $g_\textnormal{eff}$ data extracted from Fig.~\ref{fig:sfig_avoided_crossings}(b) fits to the function $g_{\text{eff}}/2\pi=11.3\textnormal{ MHz }\times \, J_1\big(  I_p/53 \,\mu\textnormal{A} \big)$, where $ I_p$ is in units of $\mu$A. The estimated coupling of $g=11.3\,\textnormal{MHz}$ is close to the value of $12.2\pm 0.1\,\textnormal{MHz}$ observed from the transmon-cavity avoided crossing. The factor $\nu$ can be calculated analytically using the formula 
\begin{equation}
    \nu = \left(\frac{2\pi}{\Phi_0}\frac{d\omega_T}{d\varphi_e}\frac{L}{\omega_m}\right)^{-1},
\end{equation}
where $\omega_T/2\pi = 6.783\,\textnormal{GHz}$ is the transmon parking frequency and $L$ is the effective mutual inductance with the SQUID. From the fitting value of $\nu = 53\,\mu\textnormal{A}$ we can calculate $L = 0.211\,\Phi_0/\textnormal{mA}$, which is considerably lower than the DC bias mutual inductance of $L_{DC} = 0.387\,\Phi_0/\textnormal{mA}$. 

\black{This lower tunability can be attributed to the flux pump not fully driving the differential mode of the SQUID loop. The spurious drive of the common mode of the SQUID is famously undesirable when implementing parametric flux drives, and the quest to achieve the pure differential drive has been actively pursued in the community recently~\cite{lu_high_2023, lu2026systematic}. The issue also manifests at large amplitudes of the parametric modulation, where we observe the full loss of contrast in our measurements, which we attribute to the qubit ionisation induced by the drive of the common mode of the SQUID. We are investigating this striking effect at the moment and we hope it will shed light on the ways to improve the asymmetry of the parametric drive for the on-chip flux line design.
In further device iterations, this can be addressed by incorporating a common-mode rejection filter on the flux line~\cite{li2026cascaded} and exploring other potential design solutions.}

We test strong parametric resonances with flux pumps at higher modulation frequency. This is shown with transmon spectroscopy experiments at different parking detunings $\Delta = 63~\textnormal{to}~232\,\textnormal{MHz}$ under a continuous flux pump with frequency $\omega_m = \Delta$. The pump creates flux sidebands at frequencies $\bar{\omega}_T+N\omega_m$, where the average transmon frequency $\bar{\omega}_T$ approximately coincides with the parking frequency $\omega_T$ for low pump amplitudes. In all spectroscopy plots (Fig.~\ref{fig:sfig_avoided_crossings}(c)), the rightmost feature corresponds to the $N=1$ sideband, which coincides with the cavity frequency and presents a frequency split corresponding to their avoided crossing. The next feature to the left corresponds to the transmon average frequency ($N=0$), and the lower $N=-1$ and $N=-2$ sidebands can be seen within range of the spectroscopy for some DC biases. All features display the same frequency splitting indicating the strong coupling of the $N=1$ sideband to the cavity for pump frequencies as high as $232\,\textnormal{MHz}$.

\black{The parametric flux modulation technique can prove useful for the resonant control in the following ways. First, the ancilla errors during resonant operation can be additionally mitigated by engineering the AC sweetspots at the cavity frequency. Next, as we have shown in this section, flux modulation allows realisation of resonant dynamics parametrically even when the actual transmon frequency is detuned from the cavity mode. Hence, resonant control can be realised with the ancilla DC-biased in its frequency sweetspot, offering enhanced transmon coherence, while retaining more than half of the bare interaction strength at the maximum of $g_\textnormal{eff}\,(I_p)$. Finally, pumping the flux-tunable ancilla at the frequency $2g\sqrt{N}$ can directly drive the within-manifold transition $\ket{N\,-}\leftrightarrow \ket{N\,+}$, extending the gate set of the resonant control protocol. These ideas constitute an interesting future study in the direction of further enhancing the quality of the resonant control scheme.}

\section{Quantum state reconstruction}\label{ssec:quantum_state_recon}

Here, we explain the procedure for reconstructing a quantum state, represented by its density matrix, with truncation dimension $D$, given the Wigner measurement data. An arbitrary $D$ dimensional quantum state $\rho$ is fully characterised by its $D^2-1$ independent real parameters, which we write in a vector form $\vec Y$. There are many choices for the parametrisation of $\rho$. We take the first $D-1$ diagonal elements of $\rho$ as well as the $D^2-D$ real and imaginary off-diagonal (upper triangular) elements. This way, the vector $\vec Y$ consists of real elements, which fully characterise the density matrix $\rho$.

On the other hand, the Wigner measurement that we performed is expressed as $W(\alpha)=\text{tr}(\hat P \hat D(\alpha)^{\dagger} \rho\hat D(\alpha))$ with $\hat P=e^{i\hat a^{\dagger}\hat a\pi}$ being the parity operator and grid-based displacement points $\{\alpha\}$. Let us express all the Wigner data in a vector form $\vec X$. Consequently, the input state and Wigner data are related linearly, which can be written in a single matrix equation $\vec X = M\vec Y +\vec V$, where $M$ and $\vec V$ denote a mapping matrix and a constant vector, which can easily be computed given the choice of displacements $\{\alpha\}$.

The state reconstruction is carried out in a few steps. First, we perform linear inversion (LI) to get the estimated state parameters $\vec Y_{\text{est}}$, given the Wigner measurement data $\vec X$: $\vec Y_{\text{est}}=M^+ (\vec X-\vec V)$, where $M^+=(M^{\dagger}M)^{-1}M^{\dagger}$ is the left Moore-Penrose pseudoinverse of the matrix $M$. Then, we construct the density matrix $\rho_{\text{LI}}$ based on the estimated parameters. As errors are present in experiments, this density matrix might not be physical, i.e., it might contain negative eigenvalues. Thus, given $\rho_{\text{LI}}$, we employ Bayesian inference (details below) to obtain a posterior probability distribution that accurately represents experimental data. From this distribution, we can sample density matrices and compute the mean fidelity and its standard deviation. 

\subsection{Bayesian inference}
A popular technique to obtain a physical density matrix from experimental data in the presence of noise is maximum likelihood (ML), which gives us the closest physical density matrix $\rho_{\text{ML}}$ from the obtained density matrix $\rho_{\text{LI}}$ via linear inversion~\cite{smolin2012efficient}. However, it is statistically less accurate. For example, it often leads to $\rho_{\text{ML}}$ having zero eigenvalues~\cite{blume2010optimal, knips2015long}. This is an issue as zero eigenvalues are interpreted as zero probabilities if we were to measure in the corresponding basis, which is not a valid conclusion as we perform finite measurement repetitions.

To address this issue, we use Bayesian inference, which is a statistically more accurate approach. Through Bayes' rule, it gives us a posterior distribution, explicitly showing uncertainty, given the measured experimental data ($\rho_{\text{LI}}$). The posterior distribution allows us to sample any functions of $\rho$ (including itself). For exact implementation of Bayesian inference, we follow an efficient protocol proposed in Ref.~\cite{lukens2020practical}. When implementing the protocol, we use the following parameters: $\alpha=1$, which establishes a uniform prior as we do not assume initial preference for the density matrix; the variance of the pseudo-likelihood $\sigma=1/N$ with $N=\#\:\cdot \: (D^2-1) $, where $\#$ is the number of experimental repetition; we also generate $2^{10}$ Markov Chain Monte Carlo samples and apply a thinning parameter of $2^7$ to mitigate correlations within the sample chain. 

For example, we use this procedure to sample a large number of possible physical density matrices $\{\rho_{\text{BI},i}\}$ and compute their fidelities $F_i=(\text{tr}(\sqrt{\sqrt{\rho_{\text{BI},i}}\:\rho_{\text{tar}}\: \sqrt{\rho_{\text{BI},i}} }))^2$ to a target state $\rho_{\text{tar}}$. The fidelity and its error presented in the main text are the average fidelity and its standard deviation.


\section{Protocol simulation and error budget}\label{ssec:fidelity_and_error}

We estimate the effect of different error sources over the resonant control protocol by solving the master equation of the system. 

First, we simulate the state preparation of states $|\psi_1\rangle = \frac{1}{\sqrt{2}}\left(|0\rangle + | 3\rangle\right)$ and $|\psi_2\rangle = \tfrac{1}{2}|0\rangle + \tfrac{\sqrt{3}}{\sqrt{8}}\left( i|2\rangle+ |4\rangle \right)$ using sequential gaussian pulses for each transition. The system Hamiltonian was reconstructed using the experimentally calibrated transitions from Tab.~\ref{tab:cal_sidebands}. The cavity lifetime $T_{1,\textnormal{cav}}$, transmon lifetime $T_1$ and pure dephasing $T_{\phi}= \left(\frac{1}{T_2} - \frac{1}{2T_1}\right)^{-1}$ are implemented with the respective jump operators. The adiabatic decoupling is taken as ideal and implemented with a change of basis operation. The Wigner tomography executed in the dispersive regime is also reproduced in simulation with experimental parameters, namely a dispersive shift $\chi/2\pi = 1.62\,\textnormal{MHz}$, a parity-mapping time of $272\,\textnormal{ns}$ and gaussian $\pi/2$ pulses with $\sigma_{\pi/2}$ of $8\,\textnormal{ns}$ and total duration of $4\sigma$. The simulated tomography is then rescaled following the amplitude of the Wigner function of vacuum obtained with consistent simulation parameters. The state fidelity is then estimated from density matrix reconstruction following the procedure described in Sec.~\ref{ssec:quantum_state_recon}.

\begin{figure}[h!]
    \centering
    \includegraphics[width=\linewidth]{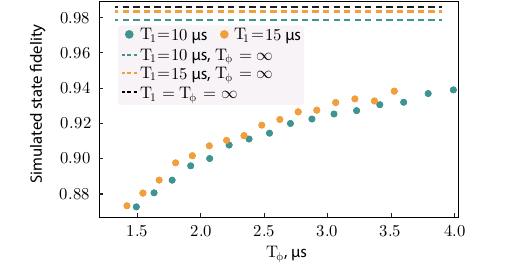}
    \caption[Simulated state preparation fidelities]{\textbf{Simulated state preparation fidelities.} Simulated state fidelity of $|\psi_1\rangle = \frac{1}{\sqrt{2}}\left(|0\rangle + |3\rangle\right)$ as a function of transmon $T_1$ and $T_{\phi}$.}
    \label{sfig:sim_stateprep_fids}
\end{figure}

The state preparation is simulated for transmon $T_1 = 10\,\mu\textnormal{s}$ and $15\,\mu\textnormal{s}$, and $T_2$ in the $\left[1.3, 2.8\right]\,\mu\textnormal{s}$ range. The cavity lifetime is fixed at $T_{1,\textnormal{cav}} = 500\,\mu\textnormal{s}$. We first identify that the transmon is highly sensitive to changes in dephasing, as seen from the reconstructed fidelity as function of $T_{\phi}$ in Fig.~\ref{sfig:sim_stateprep_fids}: the fidelity of $|\psi_1\rangle$ varies in a large range $[87.2\,\%, 93.8\,\% ]$ ($[87.3\,\%, 93.8\,\% ]$) for $T_1 = 10\,\mu\textnormal{s}$ ($15\,\mu\textnormal{s}$). These values also show the protocol is weakly sensitive to variations in transmon $T_1$ in this parameter range. Similar conclusions about $T_1$ and $T_2$ sensitivity can be drawn for state $|\psi_2\rangle$. Simulations with $T_2=2.7\,\mu\textnormal{s}$ and $T_1 = 15\,\mu\textnormal{s}$ ($T_{\phi} \approx 2.97\,\mu\textnormal{s}$) lead to $93.2\,\%$ and $87.7\,\%$ for states $|\psi_1\rangle$ and $|\psi_2\rangle$, respectively, reproducing well the experimental fidelities of  $93\pm3\,\%$ and $89\pm2\,\%$.

We calculate the fidelities of $|\psi_1\rangle$ and $|\psi_2\rangle$ while progressively including more sources of error in the simulation (Tab.~\ref{tab:state_prep_error}). The results indicate that the dephasing $T_{\phi}$ has the largest impact on state preparation fidelity, followed by coherent errors due to finite pulse length, and transmon $T_1$, in that order. From these simulations, we expect that the quality of state preparation can be significantly enhanced even with modest improvements in the transmon $T_\phi$.

We simulate the Givens rotations between Fock states $|1\rangle$ and $|4\rangle$ using the same procedure, with the addition of the optimised $|1+\rangle \leftrightarrow |3-\rangle$ state transfer drive. The simulated cavity photon distributions are compared with the experimental reconstructed density matrix in Fig.~4c in the main text, showing a close match. 

Simulations indicate that the experimental state fidelities after Givens rotations are significantly affected by coherent errors. When neglecting the transmon $T_1$ and $T_2$, the state transfer $|1+\rangle \to |3-\rangle$ shows a fidelity of $95.0\,\%$ when the system is initialised in $|1+\rangle$. For the same drive, the returning $|3-\rangle \to |1+\rangle$ path shows a fidelity of $85.8\,\%$ when the system is initialised in $|3-\rangle$. We attribute these imperfections to the mismatch between the ideal Jaynes-Cummings model used in the drive optimization and the experimental Hamiltonian parameters, which leads to a difference on the order of $\approx 1\,\textnormal{MHz}$ in the frequencies of higher transitions. We expect that further improvements in our calibration and optimization methods will render significant improvement in the Givens rotations fidelities.

\begin{table}[h]
\centering
\begin{tabular}{|c|c|c|}
\hline
Included error sources &  $|\psi_1\rangle$ fid. (\%) & $\, |\psi_2\rangle$  fid. (\%) \,  \\
\hline \hline
Finite pulse length & $98.4$ & $95.1$  \\
 + $T_1$ & $98.3$ & $94.4$  \\
 + $T_{\phi}$ & $93.2$ & $87.7$  \\
\bottomrule
\addlinespace
\end{tabular}
\caption[T]{\textbf{Error budget for state preparation.} State fidelity for $|\psi_1\rangle =\frac{1}{\sqrt{2}}\left(|0\rangle + |3\rangle\right)$ and $|\psi_2\rangle = \tfrac{1}{2}|0\rangle + \tfrac{\sqrt{3}}{\sqrt{8}}\left( i|2\rangle+ |4\rangle \right)$ states with progressive addition of error sources.}
\label{tab:state_prep_error}
\end{table}

\black{Importantly, the analysis of this section is restricted to the qubit (two-level) approximation of an ancillary transmon. Furthermore, the adiabatic decoupling ramp is replaced by an instantaneous change of basis. We investigate the impact of theses factors in the following sections.}

\subsection{Finite anharmonicity}
\black{The Jaynes-Cummings Hamiltonian on resonance is solvable analytically for the case of the two-level qubit. However, once a real transmon qubit with finite anharmonicity $\alpha/2\pi$ is considered, the JC ladder energy levels represent an approximate solution in the limit of $\alpha \gg g$ \black{(which is the case for the current device)}. To account for this, we define transmon as a four-level system with the anharmonicity $\alpha/2\pi \approx$ $-$206 MHz and diagonalise the Jaynes-Cummings Hamiltonian. We find that there are addional levels in the eigenenergies spectrum due to the higher excited transmon states, as summarised in the Table \ref{table:anharmonic_eigenenergies}. We label those non-computational levels as $\ket{\mathcal{E}_m}$ where $m$ indicates the index in the order of increasing eigenenergy.}

\black{Table \ref{table:anharmonic_eigenenergies} shows that the numerical diagonalisation of a Jaynes-Cummings Hamiltonian with the parameters given in the Table \ref{tab_supple:device_parameters} matches the experimentally measured eigenenergies within 0.002\%. Those parameters are used for numerical simulations, if not explicitly stated otherwise.}

\black{Having included anharmonicity in our model, the pulse-level simulation of the experimental state preparation protocol for $\tfrac{1}{\sqrt{2}}\left(|0\rangle + |3\rangle\right)$ shows that without decoherence the fidelity of 0.968 to the target state can be reached using the pulse durations and amplitudes specified in the Table \ref{tab:cal_sidebands}. We also found that optimising the pulse durations and amplitudes numerically for the specific state preparation protocol while explicitly considering the anharmonic system can reach fidelity $>$0.995 in the same state preparation time. }

\begin{table*}[t]
    \begin{center}
    \begin{tabular}{||c| c| c||} 
     \hline
     State label & $E_\textnormal{exp}$, GHz & $E_\textnormal{transmon}$, GHz \\ [0.5ex] 
     \hline
     $\ket{0}$ & 0 & 0 \\ 
     \hline
     $\ket{1\, -}$ & 6.8566 & 6.8566 \\
     \hline
     $\ket{1\, +}$ & 6.8809 & 6.8808\\
     \hline
     $\ket{\mathcal{E}_1}$ & $--$ & 13.5300 \\
     \hline
     $\ket{2\, -}$ & 13.7211 & 13.7211\\
     \hline
     $\ket{2\, +}$ & 13.7553 & 13.7552\\
     \hline
     $\ket{\mathcal{E}_2}$ & $--$ & 19.9871 \\
     \hline
     $\ket{\mathcal{E}_3}$ & $--$ & 20.3984 \\
     \hline
     $\ket{3\, -}$ & 20.5868 & 20.5868 \\
     \hline
     $\ket{3\, +}$ & 20.6285 & 20.6285\\
     \hline
     $\ket{\mathcal{E}_4}$ & $--$ & 26.8548 \\
     \hline
     $\ket{\mathcal{E}_5}$ & $--$ & 27.2668 \\
     \hline
     $\ket{4\, -}$ & 27.4531 & 27.4531 \\
     \hline
     $\ket{4\, +}$ & 27.5010 & 27.5011\\
     \hline
     $\ket{\mathcal{E}_6}$ & $--$ & 33.7224 \\
     \hline
     $\ket{\mathcal{E}_7}$ & $--$ & 34.1352 \\
     \hline
     $\ket{5\, -}$ & $--$ & 34.3197 \\
     \hline
     $\ket{5\, +}$ & 34.3732 & 34.3733\\
     \hline
     $\ket{\mathcal{E}_8}$ & $--$ & 40.5901 \\
     \hline
     $\ket{\mathcal{E}_9}$ & $--$ & 41.0035 \\
     \hline
     $\ket{6\, -}$ & 41.1875 & 41.1866 \\
     \hline
     $\ket{6\, +}$ & $--$ & 41.2453\\
     \hline
     $\ket{\mathcal{E}_{10}}$ & $--$ & 47.4578 \\
     \hline
     $\ket{\mathcal{E}_{11}}$ & $--$ & 47.8719 \\
     \hline
     $\ket{7\, -}$ & $--$ & 48.0538 \\
     \hline
     $\ket{7\, +}$ & 48.1166 & 48.1170\\
     \hline
    \end{tabular}
    \caption{\black{Eigenenergies in the units of GHz of the Jaynes-Cummings Hamiltonian on resonance determined from the experiment $E_\textnormal{exp}$ and from a numerical Hamiltonian diagonalisation for an anharmonic transmon with the parameters listed in text.}}
    \label{table:anharmonic_eigenenergies}
    \end{center}
\end{table*}

\black{We compare the state preparation fidelity for the same system lifetimes for the experimental pulse sequence for the case of the two-level qubit and anharmonic four-level transmon. We observe that the simulation for the two-level qubit yields $\approx$3\% higher fidelity than in the case of four-level anharmonic transmon.  For example, for qubit $T_1$ = 8 $\mu$s, $T_2$ = 3.8 $\mu$s, cavity $T_1$ = 500 $\mu$s, qubit simulation yields 0.957 fidelity, while considering anharmonic transmon leads to fidelity of 0.929.}

\black{We argue that this decrease in fidelity can be avoided by simple optimisation of pulse durations and amplitudes for the given state preparation task. For example, the system with the same decoherence values and the anharmonic transmon yields fidelity to the target state of 0.955 once the pulses are optimised. Thus, using more careful optimisation avoids the fidelity drop due to the finite anharmonicity transmon.}

\subsection{Adiabatic detuning}\label{sec_adiabaticdetuning}
\black{
Consider an ideal general state of the system before the adiabatic detuning process:
\begin{eqnarray}
    |\psi\rangle&=&\alpha|M+\rangle + \beta|(M-1)+\rangle + \gamma|(M-2)+\rangle + \cdots \nonumber \\
    &=&\frac{1}{\sqrt{2}}(|g\rangle |\psi_g\rangle + |e\rangle |\psi_e\rangle),\label{EQ_psiAD}
\end{eqnarray}
where $|\psi_g\rangle=\alpha|M\rangle + \beta |M-1\rangle+\gamma|M-2\rangle + \cdots$ and $|\psi_e\rangle=\alpha|M-1\rangle + \beta |M-2\rangle+\gamma|M-3\rangle + \cdots$ are normalised states of the cavity.
Under transmon energy decay, the transmon state transformation in Eq.~(\ref{EQ_QT1}) applied to the density matrix version of Eq.~(\ref{EQ_psiAD}) yields
\begin{eqnarray}
    \rho &=&\frac{1}{2}\big( |g\rangle \langle g||\psi_g\rangle \langle \psi_g| + |g\rangle \langle e||\psi_g\rangle \langle \psi_e| e^{-\frac{T}{2T_1}} \nonumber \\
    &&+ |e\rangle \langle g||\psi_e\rangle \langle \psi_g| e^{-\frac{T}{2T_1}} + |e\rangle \langle e||\psi_e\rangle \langle \psi_e|e^{-\frac{T}{T_1}} \nonumber \\
    &&+ |g\rangle \langle g||\psi_e\rangle \langle \psi_e|(1-e^{-\frac{T}{T_1}}).
\end{eqnarray}
The fidelity then reads
\begin{eqnarray}
    F&=&\langle \psi|\rho |\psi\rangle \nonumber \\
    &=&1-\frac{T}{2T_1}+\frac{|\langle \psi_g|\psi_e\rangle|^2T}{4T_1}.
\end{eqnarray}
The infidelity is upperbounded as $\epsilon_1=T/2T_1-|\langle \psi_g|\psi_e\rangle|^2T/4T_1\le T/2T_1$.
Similar analysis considering only transmon dephasing gives infidelity $\epsilon_2=T/2T_{\phi}$.
Therefore, the total fidelity can be written as $F=1-\epsilon$, where the total infidelity is $\epsilon=\epsilon_1+\epsilon_2\lesssim T/(2T_1)+T/(2T_{\phi})\approx 0.06$, which should be interpreted as an upper bound of the error.
}

\black{We performed full numerical simulation to verify that our claims and understanding are correct. We simulate the $\tfrac{1}{\sqrt{2}}\left(|0\rangle + |3\rangle\right)$ state preparation on the pulse-level using the experimentally calibrated pulse amplitudes and durations for the system with the transmon decoherence rates in range of $T_1 \in [$10, 15$]$ $\mu$s and $T_2 \in [$1.3, 2.8$]$ $\mu$s, and cavity $T_{1, \textnormal{cav}}$ = 500 $\mu$s. 
We use Jaynes-Cummings Hamiltonian in the lab frame with the finite anharmonicity transmon. We provide the upper bound on the infidelity related to the adiabiatic detuning by adding 200\,ns idling time \textit{on resonance} after the state preparation is done. Again, this is a simplified model that should only serve to provide an upper bound on the induced error.}

\black{Under these conditions, we find that adding an additional 200~ns on top of the 640~ns sequence used to prepare $\tfrac{1}{\sqrt{2}}\left(|0\rangle + |3\rangle\right)$ leads to the decrease in the state preparation fidelity from 0.862-0.916 to 0.814-0.890. This reduction on the fidelity by 3 to 5\% is consistent with the upper bound our naive analytical model predicted. Importantly, the impact of the adiabatic detuning of the final fidelity decreases with longer system coherence times, so its effect will be mitigated with more careful device design and the fabrication process.}

\black{Intuitively, we expect the actual infidelity incurred by this is a fraction of this upper bound as the transmon state is not fully hybridized with the cavity throughout the entirety of the 200\,ns. 
This value will be further reduced in future refinements of the device and protocol, where $g$ can be increased further (allowing for shorter duration of adiabatic detuning) and more sophisticated pulse shapes can be adopted to minimize the time needed for this decoupling step.}

\black{We would like to highlight here that the capability of implementing the adiabatic detuning is important for the flux-tunable device discussed in this work. Using various values of the detuning $\Delta$, we can bias the ancillary transmon in both strong and weak dispersive regime, providing access to Wigner and characteristic function tomographies in the same device. Moreover, we can suppress the cavity self-Kerr and its stray dispersive coupling to the qubit for the bosonic quantum memory applications.}

\subsection{Characterizing Givens rotation}

\black{
In the main text, we show the application of Givens rotations for a few angles $\theta=0,\,\pi/4,\,\pi/2$, and $\pi$, where the fidelities of the corresponding final state are $0.80,\,0.76,\,0.71$, and $0.68$, respectively. 
This process is well captured in our simulations with decoherence, where the fidelities are $0.80,\,0.78,\,0.72$, and $0.63$, respectively.
To illustrate gate characterization further, let us take an example of the $\hat{\mathcal{R}}_{14}(\pi/4)$ gate.
A simple characterization of this gate can be performed in the subspace spanned by $\{|1\rangle,|4\rangle\}$, which requires $4$ different input states via process tomography or application of many instances of random Clifford gates and their inverse via a more rigorous randomized benchmarking, the data of which we do not currently have.
With the help of simulations, we prepare four different input states $|1+\rangle$, $|4+\rangle$, $(|1+\rangle + |4+\rangle)/\sqrt{2}$, and $(|1+\rangle + i|4+\rangle)/\sqrt{2}$ (tomographically complete set); perform the $\hat{\mathcal{R}}_{14}(\pi/4)$ gate for each input; and record the corresponding output state.
From the set of input and output states, the average state fidelity of the four instances is $F_{\text{ave}}\approx 0.79$.
We proceed to reconstruct the Choi matrix $\Lambda$ and standardly compute the process fidelity as $F_{\text{process}}=(\text{tr}(\sqrt{\sqrt{\Lambda_{\text{id}}}\Lambda\sqrt{\Lambda_{\text{id}}}}))^2\approx 0.78$ and average gate fidelity as $F_{\text{ave}}=(2F_{\text{process}}+1)/3\approx 0.85$, which is interpreted as the average state fidelity over arbitrary input states.
Increasing the qubit coherences to $T_1=100\,\mu$s and $T_{\phi}=50\,\mu$s would result in $F_{\text{process}}\approx 0.89$ and $F_{\text{ave}}\approx 0.92$.
}

\subsection{Scaling with larger coupling g}

\black{The main advantage of the all-resonant control comes into play with larger values of the Jaynes-Cummings coupling $g$. Our full numerical simulations demonstrate that even for the modest value of $g/2\pi$ = 50 MHz the target state $\tfrac{1}{\sqrt{2}}\left(|0\rangle + |3\rangle\right)$ can be prepared in mere 168 ns.
To account for adiabatic detuning, we include an additional 50~ns of on-resonance system idling, noting that this time also scales with $g$. Across the same range of decoherence values mentioned in \ref{sec_adiabaticdetuning}, this yields a lower bound on fidelity between 0.921 and 0.953.
According to our simulations, such a performance is beyond what GRAPE can achieve. }

\black{The performance can be improved further with even higher values of $g$ up to where RWA stops working, which we estimate to be $g/2\pi\sim$ 690 MHz for the cavity frequency considered here.} \black{Note that the ancilla anharmonicity needs to be enhanced accordingly.}

\section*{Data and code availability}


\black{All numerical simulation scripts and data analysis routines — including those for quantum state reconstruction and Bayesian inference — are publicly accessible on GitHub, \href{https://github.com/Qcrew/Valadares-Dorogov-FAR}{github.com/Qcrew/Valadares-Dorogov-FAR}. The repository also contains the averaged experimental datasets supporting this study.}

\newpage
\bibliography{references.bib}